\documentstyle[sprocl,psfig]{article}

\def\be{\begin{equation}}
\def\ee{\end{equation}}
\def\bea{\begin{eqnarray}}
\def\eea{\end{eqnarray}}

\def\IR{{\hbox{{\rm I}\kern-.2em\hbox{\rm R}}}}
\def\IB{{\hbox{{\rm I}\kern-.2em\hbox{\rm B}}}}
\def\IN{{\hbox{{\rm I}\kern-.2em\hbox{\rm N}}}}
\def\IC{\,\,{\hbox{{\rm I}\kern-.59em\hbox{\bf C}}}}
\def\IZ{{\hbox{{\rm Z}\kern-.4em\hbox{\rm Z}}}}
\def\IP{{\hbox{{\rm I}\kern-.2em\hbox{\rm P}}}}
\def\IH{{\hbox{{\rm I}\kern-.4em\hbox{\rm H}}}}
\def\ID{{\hbox{{\rm I}\kern-.2em\hbox{\rm D}}}}
\def\II{{\hbox{\rm I}\kern-.2em\hbox{\rm I}}}

\def\Q{{\cal Q}}
\def\e#1{e^{#1}}

\begin{document}
\title{{\'E}tudes on D--Branes}
\author{Clifford V. Johnson\\\bigskip}
\address{Department of Physics and Astronomy\\
177 Chem--Phys Building\\
Lexington, KY 40506--0055, USA\\
\smallskip{\it cvj@pa.uky.edu}}
%
%

\maketitle \abstracts{This is a short arrangement of notes on
D--branes, offered as an embellishment of five lectures which were
presented at the 1998 Trieste Spring School entitled
``Non--Perturbative Aspects of String Theory and Supersymmetric Gauge
Theory''.  There is a good number of collections of pedagogical notes
on D--branes in the literature, and since space here is limited, no
attempt will be made to cover all of the introductory material again.
Instead, the notes cover selected topics and themes in string theory
and M--theory, emphasizing how certain technical aspects of D--branes
play a role.  The subject is developed mainly from the perspective of
non--perturbative string theory, touching on aspects of the old matrix
model, string duality, the new matrix model, the AdS/CFT
correspondence and other gauge theory/geometry correspondences.}

\medskip
\noindent
{\bf Contents}
\medskip

\contentsline {section}{\numberline {1}{\it Prelude:} Remarks on Ten Years of 
Non--Perturbative Strings}{}
\contentsline {subsection}{\numberline {1.1}\sl The First Wave: 1988/1989 and Beyond}{3}
\contentsline {subsection}{\numberline {1.2}\sl The Second Wave: 1994/1995 and Beyond}{9}
\contentsline {subsection}{\numberline {1.3}\sl Cadenza: Beyond Large $N$?}{12}
\contentsline {section}{\numberline {2}{\it Fugue:} $SO(32)$ String Duality and 
the Role of Extended Objects\hskip-0.5cm}{}
\contentsline {subsection}{\numberline {2.1}\sl Dual Actions}{14}
\contentsline {subsection}{\numberline {2.2}\sl The Logic of Duality}{16}
\contentsline {subsection}{\numberline {2.3}\sl Dual Strings}{18}
\contentsline {subsection}{\numberline {2.4}\sl Collective Motions and World Volume 
Theories}{20}
\contentsline {subsection}{\numberline {2.5}\sl Dual Five--Branes}{23}
\contentsline {subsection}{\numberline {2.6}\sl More Branes From The Other Extended 
Algebras}{25}
\contentsline {section}{\numberline {3}{\it Trio:} From $p$--Branes to D$p$--Branes}{}
\contentsline {subsection}{\numberline {3.1}\sl Trouble at the Core?}{26}
\contentsline {subsection}{\numberline {3.2}\sl Clues From Anomalies}{29}
\contentsline {subsection}{\numberline {3.3}\sl An Economical Resolution}{31}
\contentsline {section}{\numberline {4}Type\nobreakspace {}I String Theory Under
 the Microscope: Dual Strings\hskip-0.5cm}{}
\contentsline {subsection}{\numberline {4.1}\sl D9--Branes}{32}
\contentsline {subsection}{\numberline {4.2}\sl D9--Branes and D1--Branes: The Dual Heterotic
 String}{32}
\contentsline {subsection}{\numberline {4.3}\sl From D1--branes to Fundamental Strings}{36}
\contentsline {subsection}{\numberline {4.4}\sl Cadenza: Where Is The Fundamental 
Type\nobreakspace {}I String?}{39}
\contentsline {section}{\numberline {5}Type\nobreakspace {}I String Theory Under 
the Microscope: Instanton Redux\hskip-0.5cm}{}
\contentsline {subsection}{\numberline {5.1}\sl D9--Branes and D5--Branes}{41}
\contentsline {subsection}{\numberline {5.2}\sl Cadenza: ADHM Gauge Theory as String Theory on a Throat}{44}
\contentsline {section}{\numberline {6}Recapitulation}{47}

\bigskip
\bigskip

\section{{\it Prelude:} Remarks on Ten Years of Non--Perturbative Strings}
The last ten years have seen remarkable progress in our understanding
of the physics of non--perturbative string theory. Some may be
surprised that this presentation goes
 as far back as 1988, but it is
not without justification, as will be clear shortly.

The first wave of progress in this area was made in the area of
``non--critical'' string theory~\cite{polyakov,KPZ}. These are
string theories propagating in dimensions outside the critical
dimension of 26 (for bosonic strings) or 10 (for superstrings). Such
string theories (described in the conformal gauge) are in principle more 
complicated than ``critical''
strings, as there is no decoupling of the quantum fluctuations of the two
dimensional world--sheet metric (given
by the ``Liouville field'', $\varphi$) from the two dimensional degrees of freedom
representing the embedding of the string into spacetime (a two
dimensional ``matter'' theory~\cite{polyakov}). So in general, the
theory is a complicated system
involving two dimensional quantum gravity~\cite{review} coupled to a matter sector
of central charge~$c$. 

Perhaps the main challenge was to find techniques which could
facilitate the description of the gravity sector, which required
making sense of the path integral sum over random fluctuating
surfaces. By the mid--'80's, the technique of discretizing the
world--sheet appeared~\cite{discrete}, with the addition of matter
degrees of freedom within that approach following soon
after~\cite{matter,kazakovloops}. There was a definite convergence of
the results of the discrete approach with those of the continuum
approach~\cite{KPZ,DDK,staud}, by about 1988.  Certain important toy
models were quickly identified: For $c{<}1$ the matter theories are
the $(p,q)$ conformal ``minimal models'', using the standard notation,
with $c{=}1{-}6(p{-}q)^2/pq$.  Many results were obtained for the
case where the topology of the world sheet was that of a sphere
(string tree level), and perturbative techniques 
for developing the topological expansion were also available.
In 1989 non--perturbative information was
obtained~\cite{exact}. The now traditional explosion of papers
followed.

Although these models are drastically simplified string theories, some
of the lessons of learned from them about non--perturbative string theory
are still relevant in today's approaches to string theory, both
non--perturbative and perturbative. This has been already pointed out
in the current literature to some extent (see later), but one suspects
that there are a few more important lessons from those days lying in
wait to be appreciated.

The second wave of progress (1994/1995) is what is now called the ``Second
Superstring Revolution'', in which our understanding of
non--perturbative critical superstring theory was greatly improved. 
That new knowledge is phrased in terms of the dramatic phenomenon known as
``strong/weak coupling duality'', and that collection of discoveries
 deserves
to be called a revolution, in contrast  (arguably) to the previous
non--perturbative breakthrough: In 1988--1990, the
severe simplicity of the models made it hard to see just what the
general lessons were at the time (although see ref.~\cite{shenker}).
 In this case, however, although probably
 all agree that we have not fully assimilated the meaning of
duality, there are some important conclusions which have changed at
least our qualitative expectations of what string theory really is,
while at the quantitative level, the relative importance of
certain features (like the role of extended objects) has been
considerably reevaluated. In addition, there has been
 a vast the number of applications of the duality 
results to problems  not central (as far as we know) to string theory.

Of course, none of these breakthroughs, whether they are called
revolutions or not, can happen without the appropriate tools, which
will be discussed in turn. In the case of the first wave, the really sharp
tools were actually matrix models, while in the second wave, they were
D--branes, which will ultimately lead us back to matrix models.

\subsection{\sl The First Wave: 1988/1989 and Beyond} 
\begin{figure}[ht]
\hskip1.0cm
\psfig{figure=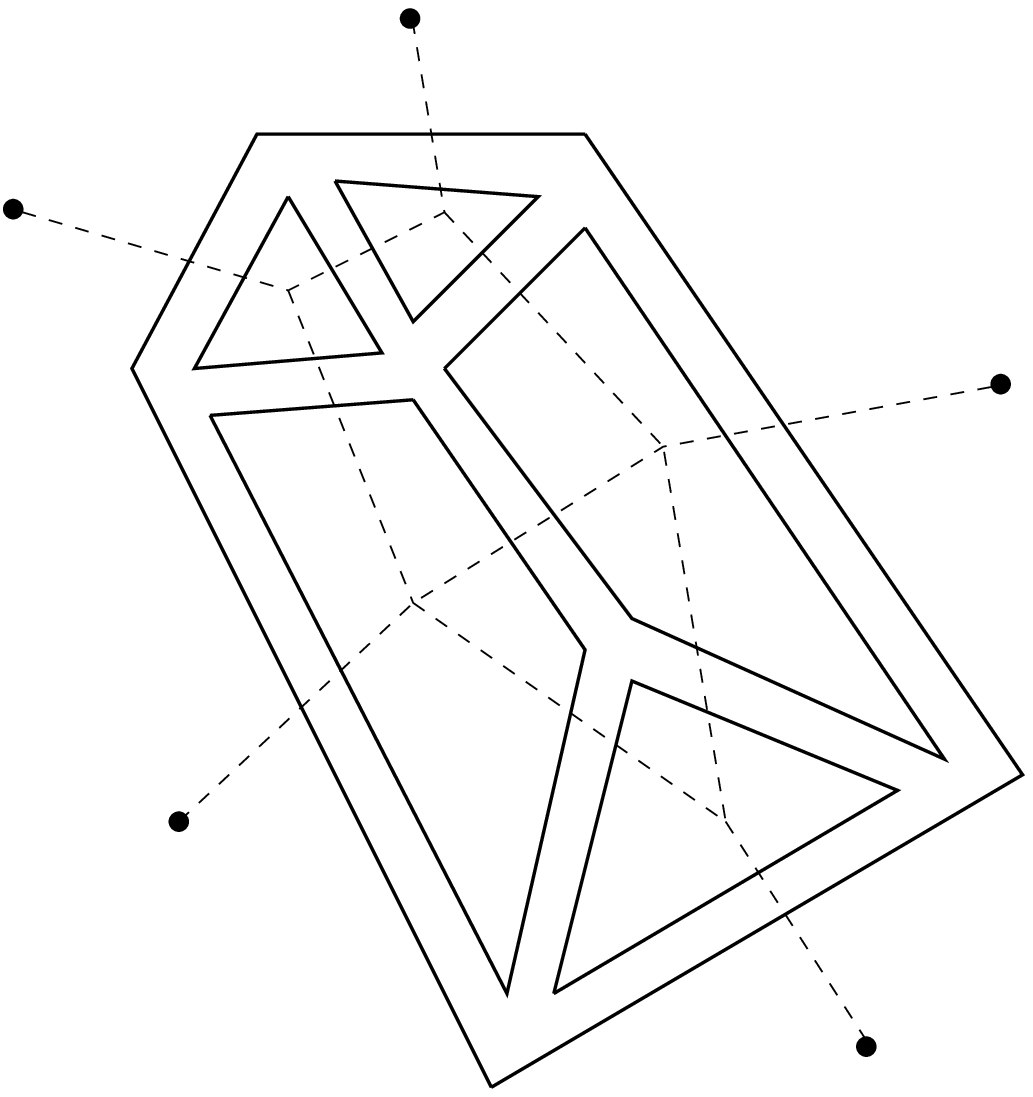,height=1.5in}
\hskip1.0cm\psfig{figure=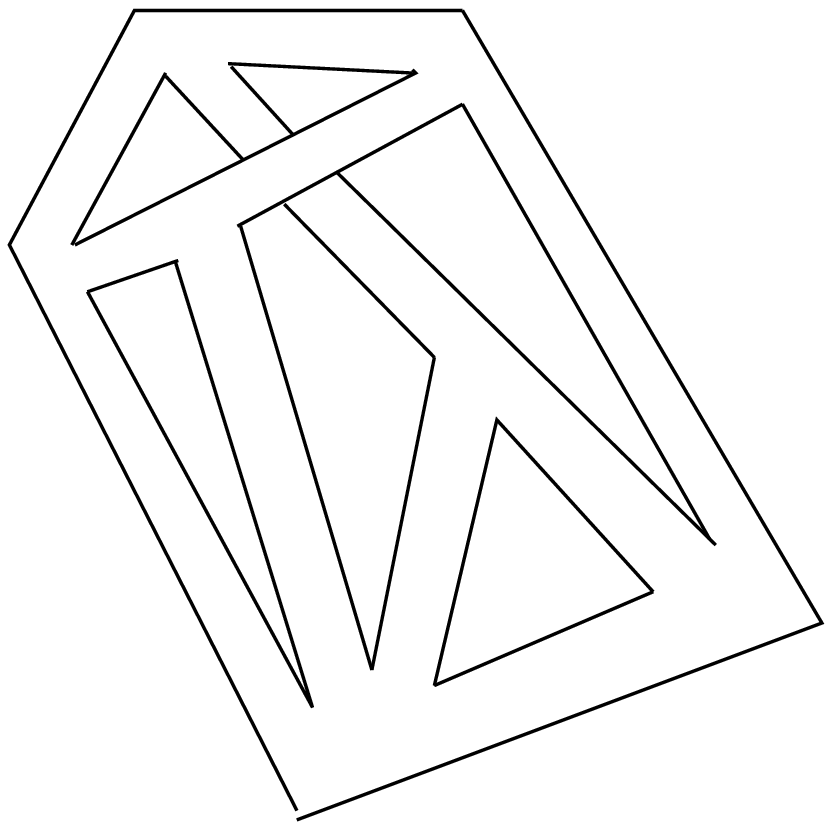,height=1.2in}
\caption{A sphere and a torus, respectively, each tiled by six
triangles and a square by the dotted dual diagram. 
The dots should be identified. The completion
of the second diagram is left for the reader's amusement.}
\label{graphs}
\end{figure}

 Matrix models allowed for a complete solution of the problem of
coupling two dimensional quantum gravity to various matter systems
with central charge~$c{\leq}1$.  This worked very well because the
Feynman graphs of certain theories of $N{\times}N$ matrices $M$ are
dual (in a sense to be made clear shortly) to ``polygonizations'' of
two dimensional surfaces. For example, for the $(2,2m{-}1)$ minimal
models coupled to gravity, the partition function is:
\begin{equation}
\label{partfun}
Z=\int dM \exp\left(-{N\over\gamma}{\rm Tr}V(M) \right)
\end{equation}
where $V(M)=\sum_{i=2}^k g_iM^i$ is a polynomial in an $N{\times}N$ 
Hermitian~\footnote{One can have  other
types of matrix ensemble: unitary, 
general complex, symmetric, {\it etc.}
Complex matrix models can be tuned to give a rich class of 
models~\cite{complex,stable} which includes 
the ones discussed here. 
Symmetric matrix models~\cite{neuberger} give unoriented strings, while unitary~\cite{periwal}
ones lead to continuum models which have been shown to be ``dual''~\cite{matrixdual} 
to certain open string models~\cite{kostovopen}.}
matrix $M$ of order~$k$ (which is at least $m$), with couplings $\{g_i\}$.  At
large $N$, one can organize the graphs, which contain vertices of up
to $k$th order, as an expansion in $1/N$. A graph of $V$ vertices, $F$
faces and $E$ edges comes with a factor $N^{F-E+V}{=}N^\chi$, where
$\chi$ is the Euler number of the graph. Replacing each vertex, face
and edge of the graph by the face, vertex and edge of dual graph, one
arrives at a polygonization of a two dimensional surface of genus $h$
where $\chi{=}2{-}2h$, and $h$ is the number of handles. For example,
in figure (\ref{graphs}), the figure on the left is dual to a tiling of
the sphere by six triangles and one square, while the figure on the
left is  dual to a tiling of the torus with the same shapes.

In this way, we see that the sum over random graphs represented by the 
integral (\ref{partfun}) is equivalent to a sum over two dimensional discretized
random surfaces.
Of course, in using this as a model of continuum string world sheets,
the physics should not depend upon whether we use triangles ($k{=}3$),
squares ($k{=}4$), or pentagons ($k{=}5$), {\it etc.,} to tile the
surface. In the continuum limit (when the size of the polygons shrinks
away) we should obtain universal results. For arbitrary couplings
$\{g_i\}$ this is not true. However, for the order $k$ potential, there
is a family of critical couplings $\{g_1^c,g_2^c,\ldots,g_k^c\}$ for
which universal physics (earmarked by certain critical
exponents~\cite{KPZ}) may be extracted, by tuning them to a place where
the free energy of the model (divided by $N^2$) becomes non--analytic.

Physically, this means that in the large $N$ limit, not only do the
graphs which can be drawn on the sphere dominate because they are an
$N^2$ factor greater than the others, but the {\it numerical coefficients}
of those graphs (which depend on the $\{g_i\}$) are such that the
perturbative expansion in the $\{g_i\}$ ceases to converge.

\begin{figure}[ht]
\hskip-0.3cm\psfig{figure=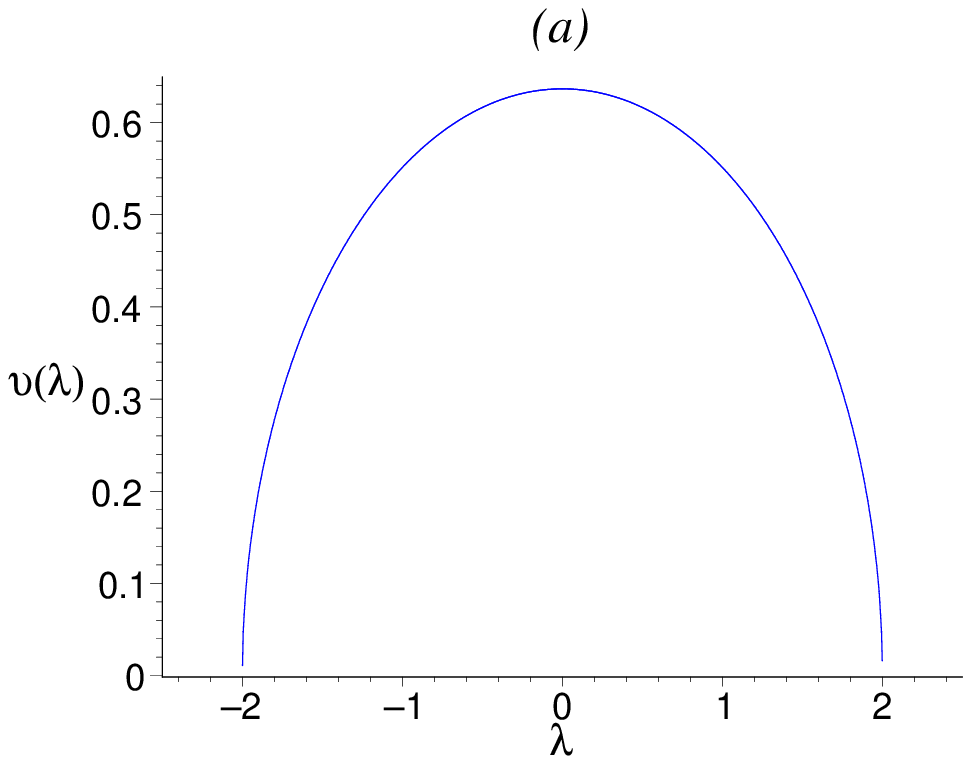,height=1.3in}
\hskip-0.3cm\psfig{figure=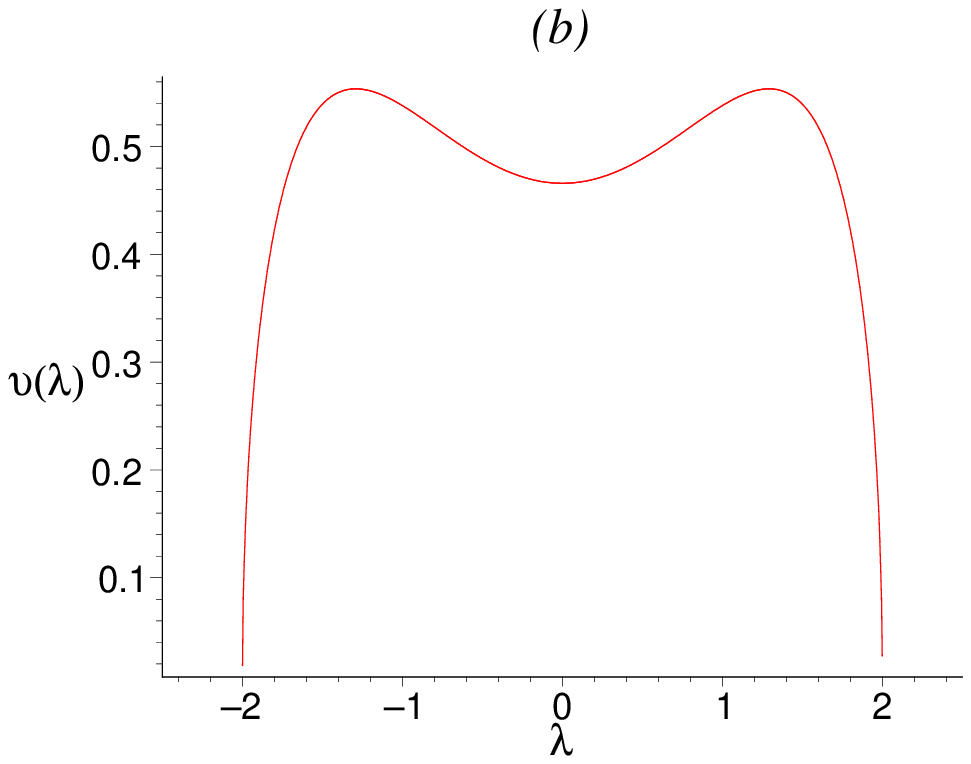,height=1.3in}
\hskip-0.3cm\psfig{figure=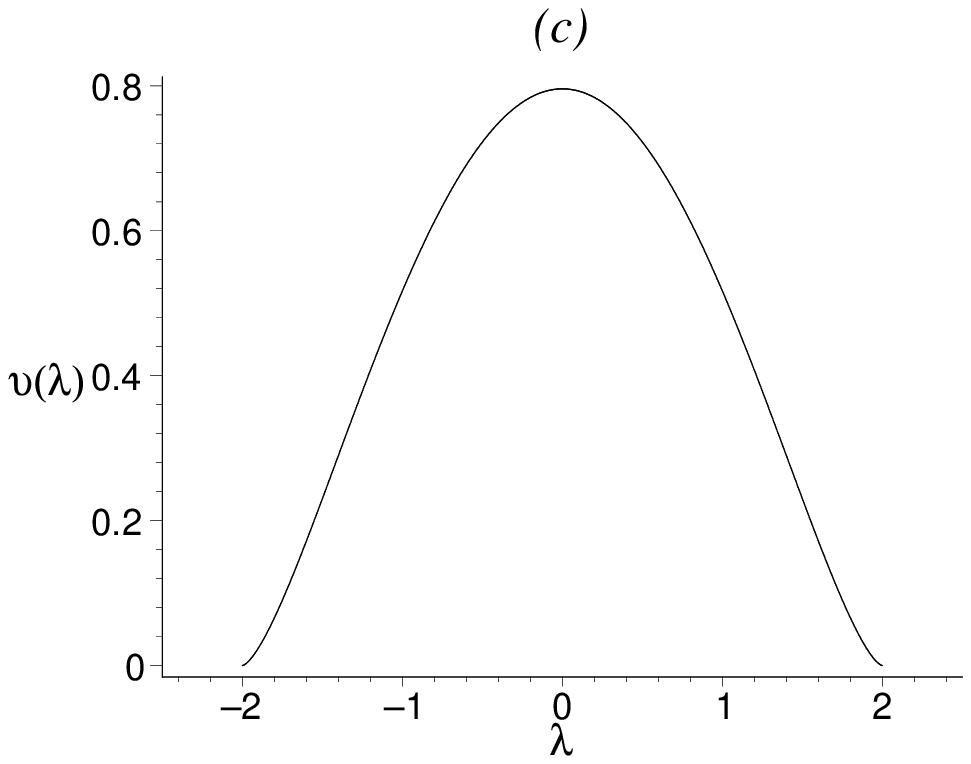,height=1.3in}
\caption{Eigenvalue densities. Case (a) is the Wigner semi--circle
bahaviour. Case (b) represents an arbitrary point in coupling space,
while (c) is the critical $m{=}2$ point,
ultimately representing the case of pure gravity. (See text.)}
\label{density}
\end{figure}

This actually corresponds to tuning the potential and eigenvalue
spectrum of the model in the following way~\cite{bipz,kazakovloops}: 
The function describing
the density of matrix eigenvalues $u(\lambda)$ (representing the ``Dyson
Gas'' of eigenvalues) is supported on a square root cut of length $2a$
in the complex $\lambda$ plane and is of the form
\begin{equation}
u(\lambda)={1\over\pi}P^{(2k-2)}(\lambda)\sqrt{a^2-\lambda^2}
\end{equation}
where $P^{(2k-2)}(\lambda)$ is a $(2k{-}2)$th order polynomial in $\lambda$ 
(derived from $V(\lambda)$), and $a$ is a
function of the couplings. (Aficionados of the old, old matrix model
days, where the distribution of large nuclei was the context, will
recognize the case of Wigner's semi--circle law for $k{=}1$. See figure
\ref{density}(a).)

When one tunes the potential to the critical point there is a family
of distinct models which are distinguished by how fast the eigenvalue
density function vanishes at the edges. (It turns out that each edge
gives a copy of a critical model~\cite{bachpetro,factorize}, 
so one can focus on one end without loss of generality). 
The $m$th critical model is characterized by the
vanishing of $u(\lambda)$ with $m$ extra zeros at the endpoint:
$u(\lambda){\sim}\lambda^{m+1/2}$. Tuning this number of zeros to be at
the edge gives the critical couplings $\{g^c_i\}$ of the potential. In
this way, one can get a family of critical continuum models in the
large $N$ limit, and solve for their behaviour on the sphere. See
figure \ref{density}(c) for the case $m{=}2$. Figure \ref{density}(b) 
is a random off--critical density distribution.

It turned out that there is a way of tuning~\cite{exact} to critical couplings
$\{g^c_i\}$ while taking the large $N$ limit, and capture the
whole genus perturbative expansion in a succinct
expression, which even contains  non--perturbative
information. This careful tuning is called the ``double scaling
limit'', where all quantities of interest are tuned to their critical
values at a rate determined by their length dimension: The amount that
they differ from their critical value is proportional to a power of
$\ell$, the typical side length of a polygon in the tiling. The limit
$N{\to}\infty$ is accompanied by $\ell{\to}0$ to recover the continuum
limit at a critical point.

One must cast the quantities of interest in a form which allows the
double scaling limit to be made explicit. An important part of the
story was the re--expression of the content of the partition function 
(\ref{partfun}) in terms of its Dyson--Schwinger equations, or equivalently in terms
of recursion relations between an infinite family of polynomials
${\cal P}_n$, orthogonal 
\cite{orthog} with respect to the measure $d\lambda\exp\{-{N\over\gamma}
V(\lambda)\}$.

In the double scaling limit, the recursion relations yield a
differential equation for (essentially) the partition
function. Alternatively, the discrete Dyson--Schwinger 
equations~\cite{discreteloops} become an
infinite series of constraints (the ``Virasoro constriants'') on all
of the correlators of the theory, which has equivalent content~\cite{loops}.

For example, in the case $m{=}2$, the differential equation is the
Painlev\'e II equation:
\begin{equation}
\label{painleve}
-{\nu^2\over3}\frac{\partial^2\rho}{\partial \mu^2}+\rho^2=\mu,
\end{equation}
where 
\begin{equation}
\rho=\nu^2{\partial^2 F\over\partial \mu^2}.
\label{suscept}
\end{equation}
$F$ is the free energy. The correctly scaling part of $\gamma$ in the limit is
$\mu$, the ``cosmological constant'' of the theory 
({\it i.e.,} it is the variable coupling to the area in the two
dimensional gravity theory). $\nu$ is a renormalized $1/N$, counting
loops in the genus expansion. 
The string coupling, $g_s$, is given by the combination
$g_s^2{=}\nu^2\mu^{-{5\over2}}.$ 

\bigskip
\hskip-0.7cm\fbox{\parbox{4.65in}{{\it Obbligato:} The cosmological constant $\mu$ has
dimensions of $({\rm length})^{-2}$ and therefore $\nu$ has dimension
$-{5\over2}$ in length units. It is the ratio
$\nu{=}\ell^{-{5\over2}}/N$ which is held finite and fixed as $N{\to}\infty$ and
$\ell{\to}0$. Therefore the dimensionless string coupling $g_s$ is given by
$g_s^2{=}\nu^2\mu^{-{5\over2}}.$ More generally, for the $m$th model,
$\nu$ has dimension ${-}(2{+}{1\over m})$ and the dimensionless string coupling
is 
$g_s^2=\nu^2\mu^{-{(2+{1\over m})}}$.
}}
\bigskip

The Painlev\'e equation (\ref{painleve}) contains the complete genus
expansion of the theory. At tree level we have $\rho{=}\mu^{1\over2}$,
while higher orders can be obtained by recursively substituting
corrections into the differential equation:
\begin{equation}
\label{genus}
\rho=\mu^{1\over2}
-{1\over24}{\nu^2\over\mu^2}-{49\over1152}{\nu^4\over\mu^{9\over2}}+\cdots
\end{equation}
One can integrate twice (and divide by $\nu^2)$ to get the free energy:
\begin{equation}
F=g_s^{-2}{4\over15}+{1\over24}\ln\mu-{7\over1440}{g_s^2} +\cdots
\end{equation}
representing the sphere, torus and double torus contribution, {\it
etc}.  

The differential equation contains more than just the complete string
perturbation series, as it has non--perturbative solutions too.
Perhaps the most important information that can be deduced from the
equation and its series solution is knowledge about the high order
behaviour of string perturbation theory, and the nature of
non--perturbative effects. It was observed~\cite{shenker} that the high
genus behaviour of the $h$th term in the series expansion is of the
form $C^{-2h}(2h)!$. (Here $C$ is not an essential (but computable)
constant.) Furthermore an examination of (for example) an attempt
to perform a Borel resummation of the series reveals that the resulting 
failure~\footnote{Note that a failure of Borel re--summability is not on its own 
a sign of a problem with perturbation theory. Consider for example the case of degenerate
double wells in ordinary quantum mechanics. The non--Borel--resummability of
the perturbation theory developed there is simply a consequence of its inability
to capture the effects of
instantons representing tunneling between the degenerate wells. This
persists to more complicated examples in quantum field theory~\cite{zinn}.}
 to resum can be expressed in terms of non--perturbative ambiguities of
the form $\exp(-C\mu^{5\over4}/\nu)=\exp(-C/g_s)$. Tracing this back
to the original matrix model definition, we see that this
translates~\cite{shenker} into $\exp(-CN)$, as $1/N$ is the string
coupling before double scaling. This is to be identified with the
exponentiated action for a single matrix eigenvalue, $\lambda$, to
tunnel from the potential well given by $V(\lambda)$ at the critical
value of the couplings which defines the model. In tunneling, it
leaves a gas of $N$ eigenvalues and hence has action of order $N$.

\noindent
That characteristic high genus behaviour, and the associated
non--perturbative effects, were correctly 
predicted~\cite{shenker} to be features of the models which will persist
to the case of critical string theories. This was borne out in the physics of
D--branes~\cite{Djoe}. Now that D--branes have brought us back
to large $N$ models again (where $N$ is the number of
D--branes) we should expect to interpret order $\exp{(-N)}$ effects as the
action of a single D--brane (see next sub--section, and also
section~4).

The string equation for the $m$th critical model is
\begin{equation}
\label{stringequation}
C_m{\cal R}_m[\rho]=\mu
\end{equation}
where the ${\cal R}_m[\rho]$ form a family of differential
polynomials~\footnote{These ``Gelfand--Dikki'' polynomials~\cite{gd}
arise naturally in expanding the diagonal part of the resolvent of the one dimensional
Schr\"odinger operator ${\cal
H}{\equiv}-\nu^2\partial^2_\mu+\rho(\mu)$.}\ in $\rho$ and its
$\mu$ derivatives (a prime denotes $\nu\partial/\partial\mu$):
\begin{equation}
{\cal R}_0=1,\quad {\cal R}_1=-\rho,\quad {\cal
R}_2=3\rho^2-\rho^{\prime\prime},\quad{\cal
R}_3=10\rho^3-10\rho\rho^{\prime\prime}
-5(\rho^\prime)^2+\rho^{\prime\prime\prime\prime},\ldots
\label{gelfdik}
\end{equation}
and the $C_m$ are numbers which normalize the coefficients of $\rho^m$
to 1 in the string equations. 

One can formally define an interpolating model~\cite{kdvloops}, flowing from
one critical model to the $k$th one, by adding the operator ${\cal
O}_k$ to the model with coupling~$t_k$. (In the original matrix model, this is
equivalent to adding the $k$th critical potential and then double scaling.)
There is a beautiful underlying organization of this flow in terms of the $k$th equation of
the KdV hierarchy~\cite{kdv} of integrable flow equations~\footnote{This extends
to the full family of the $(p,q)$ minimal models in
terms of the ``generalized'' KdV hierarchy of flows. These models may be derived 
by double scaling  two--matrix models~\cite{pqdouglas,twomatrix}}:
\begin{equation}
{\partial \rho\over\partial t_k}={\cal R}^\prime_{k+1}[\rho].
\label{flow}
\end{equation}

Note that from equations (\ref{flow}) and (\ref{gelfdik}) it is
evident that $t_0{\equiv}-\mu$, and therefore ${\cal O}_0$ is the
operator that measures area, known as the ``puncture
operator''~\footnote{The name arises because one may think of ${\cal
O}_0$ as marking a fixed point on the world--sheet. Fluctuations about this 
point then give  a measure of the average world--sheet 
area.} This identification of ${\cal O}_0$
is correct for the non--trivial unitary member of the series of models, (2,3). 
For the $m$th model,
the operator ${\cal O}_{m-2}$ is the puncture operator.

Therefore, from
(\ref{suscept}), we see that the string equation is an equation for 
the two point function $\rho\equiv<\!{\cal O}_0{\cal O}_0\!>$, while the KdV flow
equation~(\ref{flow}) is an equation relating the insertion of the operator
${\cal O}_k$, $<\!{\cal O}_k{\cal O}_0{\cal O}_0\!>$, to other 
insertions, following from the fact
that the ${\cal R}_k$
obey  a recursion relation.

This KdV organization has a number of consequences, and is equivalent
to an infinite family of constraints on the correlation functions of
the point--like operators ${\cal O}_k$ appearing in any of the
theories. These constraints form a Virasoro algebra, and may be
thought of as an expression of the Dyson--Schwinger equations of the
``microscopic loops'' (point--like operators) of the
theory~\cite{kazakovloops,davidloops,loops}.  Unfortunately that is not
a story which we can cover here, due to lack of space.

Incidentally, the even $m$ models with the string equation 
(\ref{stringequation}) defined above were shown to be problematic 
at the non--perturbative level: 
A unique exact solution to the equation with the sphere--level asymptotic
behaviour $\rho{\sim}\mu^{1/m}$ does not exist for $m$
even~\cite{unstable}. Furthermore, the flow from the unique, exact solution~\cite{oddm}\ 
for an odd $m$ model to an even $m$ model, using the
KdV evolution is unstable~\cite{unstable}. 
So the even $m$ models (including the unitary model $(3,2)$,
the case of pure gravity) were considered to be sick at
the non--perturbative level.

It is worth pointing out that while this is a perfectly acceptable and
non--disturbing conclusion (given that these are toy models), there
are other string equations closely related to those above
(eqn.(\ref{stringequation})) which may be derived from slightly
different (and no less well motivated) matrix models. That family of
models has the same behaviour on the sphere, {\it and at any order in
perturbation theory} and the same underlying KdV flow structure, but
have a unique stable solution for all $m$, and therefore might be
considered to be the correct non--perturbative completion of
the perturbative series~\cite{stable}.

\subsection{\sl The Second Wave: 1994/1995 and Beyond}

In the case of the second wave of non--perturbative discoveries, the
sharp tools are D--branes, which are extended objects 
with (at least) four extra special properties.

\begin{itemize}

\item{They have a very good description within weakly coupled type~I
and type~II critical string theory in terms of the inclusion of
Dirichlet boundary
conditions~\cite{dbranesi,dbranesii,dbranesiii,dbranesiv} into the
usual conformal field theory description. Their low energy collective
motions have a simple description in terms of gauge field 
theory~\cite{leigh,Djoe,edbound}, a
fact which is not just convenient, but very fundamental, as
we shall discuss.}

\item{They are extended sources~\cite{gojoe} of a whole family
of spacetime fields, the ``Ramond--Ramond'' (R--R) fields. This was
a crucial fact in the string duality~\cite{duality} story, because R--R
fields themselves were implicated as important role players as
the duality map mixed them with ``Neveu--Schwarz--Neveu--Schwarz''
(NS--NS) fields. Extended soliton sources of the NS--NS fields were
correspondingly mapped~\cite{isstrings}
into extended sources~\cite{pbranes} of the R--R fields, which
were noticed to be much lighter and singular objects in the theory.
Those properties were thus shown to be necessary by duality, adding to evidence
\cite{stromingerii} that
R--R extended objects play a fundamental role in string
theory.}

\item{D--branes carry the smallest possible~\cite{gojoe}  R--R charges
in the theory allowed by Dirac--Nepomechie--Tietelbiom charge
quantization~\cite{quantized}, and as solitons, their collective
motions are described at low energy by gauge 
theory~\cite{leigh,Djoe,edbound}. These facts
(combined with the BPS property, below) allowed for quantitative
checks of duality statements, and led to applications beyond the issue
of non--perturbative string theory, for example to the physics of
black holes~\cite{peet} and gauge theory~\cite{giveon}.}

\item{As massive states of the type~I and type~II strings' 
extended spacetime supersymmetry algebra in ten dimensions
(or fewer by compactification), D--branes are
Bogomol'nyi--Prasad--Sommerfeld (BPS) saturated~\cite{bps}
states~\cite{gojoe,Dgreen}. As the spectrum of such states is the same 
for all values of the coupling~\cite{edolive}, many
statements made about string theory at strong coupling may be tested by
following these objects from weak coupling, where we can calculate.}

\end{itemize}

We will uncover some of these properties and connections in later
sections, with particular attention to their role in string
duality~\footnote{For introductory reviews, and reviews with advanced
applications, see for example
refs.~\citeup{dnotes,joetasi,bachas,giveon,peet}.}.  However let us
disuss a few applications for now, in order to point out an
interesting ten year cycle.

At low energy, branes have a description of their dynamics in terms of
an effective field theory living on their world volume. In the case of
D$p$--branes, where $p$ denotes the number of spatial directions into
which it extends, the effective field theory is a gauge
theory. Indeed, when there are $N$ coincident D$p$--branes the gauge
symmetry is $U(N)$ and~\footnote{In order to highlight certain
essential parts of the properties of D--branes, the introduction of
orientifolds and the objects correspondingly known as ``O$p$--planes''
will not be discussed much in this section, or in these notes.  They
are a necessary component of a complete discussion of the string
duality story, and applications to gauge theories with $SO$ of $USp$
gauge groups, orientifold model building, {\it etc.,} but space here
is limited.  (A working title for this collection of notes 
was therefore ``Supposed Former Orientifold Junkie'').}\ the field theory is
thus a model of $N{\times}N$ matrix--valued fields: a ``matrix model''
in a very precise sense, as we shall see.

Since the relevance of D--branes to strong/weak coupling
duality was noticed~\cite{gojoe}, this convenient low
energy description has had many applications, including ones beyond
string duality for its own sake (although such a distinction is of
course blurring daily).

Two such areas are the subject (at least in part) of the lectures, at
this school, of H. Verlinde and M. Douglas, respectively. One is the
 ``matrix theory'' (and its relatives, the ``matrix string theories'')
approach~\footnote{See also the 
lectures of H.~Nicolai~\cite{nicolai} 
on membranes in eleven dimensions, and their relation to
M--theory and Matrix theory.}  to the search for a definition of M--theory, while
the other is the AdS/CFT correspondence and its generalizations. Both
were motivated and initially tested using D--branes as the basic
tools, but it is largely believed that they are examples of phenomena
(examples include ``holography''~\cite{hologram}, and a 
description~\cite{edbound,noncommute} 
of spacetime in terms of a non--commutative geometry)
which are true beyond the D--brane context.

Matrix theory~\cite{matrix}\ might most readily be called the ``new matrix
model''. It is a supersymmetric quantum mechanics of $N{\times}N$
matrices, with a very particular form for its potential, dictated by
${\cal N}{=}16$ supersymmetry. In the limit of infinite $N$,
 it manages to capture the physics of
eleven dimensional supergravity (the low energy limit of
``M--theory'') in the infinite momentum frame (often 
called the ``light cone gauge'') and in
the ``discrete light cone gauge''~\cite{mlenny,mseiberg,msen} for finite~$N$. 

The AdS/CFT correspondence~\cite{juan,gubkleb,edads}\ is also a sort of new matrix model,
but in a more subtle sense. It is founded upon
the results of many investigations relating the properties of low
energy string theory in curved spacetime backgrounds to gauge theory,
tailored to the case of large $N$. Let us pause here and reflect.

The strong statement of the AdS/CFT correspondence, in the simplest 
case of 16 supersymmetries, is that string (or M--) theory propagating on a
background composed of the product of $p{+}1$ dimensional anti
de--Sitter spacetime (AdS$_{p+2}$) and a sphere $S^{8-p}$ (or
$S^{9-p}$), is dual to a $p{+}1$ dimensional conformal field theory,
for $p{=}3$ (the type IIB string), and $p{=}2,5$ (M--theory).

Weakening the statement somewhat, the relation speaks to the low energy,
and weakly coupled limit of string (or M--) theory, which is
simply the classical ten (or eleven) dimensional supergravity limit
(type IIB in the ten dimensional case). For $p{=}3$, the conformal
field theory is the $N{\to}\infty$ limit of the $SU(N)$ ${\cal N}{=}4$
supersymmetric gauge theory living on the 3+1 dimensional world volume
of the D3--brane. The Feynman diagrams of the gauge theory may be organized
at large $N$ in terms of a genus expansion~\cite{thooft}
of discretized string world--sheets, just as we saw before.
 For $p{=}2$ and 5, the theories are the exotic
conformal field theories with 16 supercharges to be found on the world
volumes of the M2-- and M5--brane solitons~\cite{mtwo,mfive} of M--theory,
respectively. More details of how to arrive at this correspondence is
reviewed and discussed in the lectures of M. Douglas in this school.

Let us focus on the case of $p{=}3$ for now. The effective string whose
world--sheets the gauge theory Feynman diagrams were constructing was the type~IIB
string compactified on AdS$_5{\times}S^5$. The correspondence works
partly because the squared radius of the $S^5$ and the inverse
 (spacetime) cosmological
constant of the $AdS_5$ are set by the combination~\cite{juan}
$R^2{=}\alpha^\prime(4\pi g_sN)^{1/2}$, in the limit that
$\alpha^\prime{\to}0$.

So in order to make the classical supergravity limit valid,
one must keep this radius $R$ well above the Planck length, in order
to honestly ignore quantum gravity corrections. So given that we are
at weak string coupling, $g_s{<}1$, we must take $N{\to}\infty$ in
order to keep $g_sN{>>}1.$ This means that $g_s{\sim}1/N$, which
should call to mind the beginning of this section. The sphere level of
critical string perturbation theory in the limit $\alpha^\prime{\to}0$ is
classical supergravity, and it captures the large $N$ physics of the
Yang--Mills theory.

The stronger statement is that the $1/N$ and $\alpha^\prime$
corrections to the large~$N$ limit are captured by the full type~IIB
string
theory, which of course goes a long way to realizing 't Hooft's
expectations set out in 1973 concerning the large $N$ expansion of
four dimensional gauge theories~\cite{thooft}.

It is hopefully clear now that we have indeed come full circle over the
past ten years. We have now (1997/1998) a strong statement that the
sphere level in critical string perturbation theory is dual to the large
$N$ limit of a ``matrix model'', which in this case is four
dimensional Yang Mills. In 1987/1988, a zero dimensional matrix
model was understood to correspond to the sphere level of a
non--critical string theory.  In both cases, their Feynman graphs are
dual to polygonizations of the world sheet of the string theory.  Now,
the string theory is a string theory propagating on
AdS$_5{\times}S^5$ while then it was a string theory
propagating~\footnote{It was pointed out~\cite{gubkleb} that the 
string propagating on
AdS is similar to the non--critical string, where the radial coordinate in AdS
is identified with the Louiville field, $\varphi$.} in $2{-}6(p{-}q)^2/pq$ 
dimensions~\footnote{The
extra dimension in the count here is because the Liouville field $\varphi$,
 acts as an extra
embedding coordinate.}.

\subsection{Cadenza: Beyond Large $N$?}
A next step is of course the understanding of $1/N$ and
$\alpha^\prime$ corrections. Unfortunately, we start to run into
difficulties. The negative (spacetime) 
cosmological constant involved in the AdS$_5$ part of the compactification
induces $N$ units of R--R flux supported on the $S^5$ (carried by the $N$
D3--branes in the dual description). This means that to directly understand
the $1/N$ and $\alpha^\prime$ corrections to the description of
string theory propagating on this geometry, we must understand more 
about how to study type~IIB string theory
propagating in backgrounds containing non--trivial R--R fields. This is
at present an under--developed subject.

We might wonder if there might be some clues to be found to guide us
in this endeavour by turning  the calendar back to 1989, when the full
renormalized $1/N$
expansion and indeed the whole non--perturbative story was
uncovered. In particular, is there an analogous procedure to tuning
couplings in the matrix model while taking $N{\to}\infty$ which will
unlock a new box of tricks for studying non--perturbative string
theory on AdS backgrounds?

Presumably we need either the analogue of the orthogonal polynomials
(employed in the study of the zero dimensional matrix models) with
which to re--express the Yang Mills partition function, or
alternatively (and perhaps more realistically) a set of
loop/Schwinger--Dyson equations for either macroscopic loops
({\it e.g.,} Wilson loops), or microscopic loops ({\it i.e.,} point--like
operators). Once this rewriting of the content of the theory is done, perhaps in
terms of a recursion relation in $N$, we might then be able to see how
to tune the coupling constants with $N$ in a manner which allows
the resummation of the $1/N$ expansion, perhaps leading to a closed
form analogous to the elegant ``string equations'' of old.

Is it at all reasonable to expect that such a procedure might be
possible? Perhaps there is some hope. Notice
that the AdS/CFT correspondence is not really founded upon
supersymmetry or string duality in any essential way, but is of course
consistent with it. So one expects that the fact that the type IIB
superstring theory has the additional features of being  highly supersymmetric
and in possession of a large $SL(2,Z)$ strong--weak coupling self
duality (as of course is the ${\cal N}{=}4$ supersymmetric 4D Yang
Mills theory), will be an additional bonus feature which allows an
elegant mathematical solution analogous to the case of
ten years ago. Given also that the Coulomb branch of many
${\cal N}{=}2$ four dimensional theories also has an elegant
description, as shown by the results of Seiberg and Witten~\cite{seibwitt}
(and in work which followed) there is similar reason to expect that for cases
with that amount of supersymmetry, there may also be a concise and
beautiful non--perturbative dual string description waiting to be
found.

If this is a complete analogue of the non--critical case, then the
following question arises: Is it possible that in some sense we are
already frozen at criticality? In analogy with the matrix models of
ten years ago, we have solved tree level in the continuum limit, and
hence our ``matrix model'' couplings $\{g_i\}$ are already tuned to their
critical values, $\{g^c_i\}$. In order to achieve the double scaling
limit which gave the complete resummed $1/N$ expansion, the couplings
had to be temporarily taken off--criticality and tuned to their
critical values at the correct rate while taking $N{\to}\infty$.

This is something that we have not yet done in this context:
 The fact that we are already
doing correct continuum tree level string physics with the AdS/CFT
correspondence translates into the fact that we are on shell, doing conformal
field theory.  The analogue of taking the defining couplings
off--criticality therefore might necessitate finding a new off--shell
definition of string theory in order to move away from the already
fixed relationships between the couplings (here, these are
the Yang--Mills coupling, the string coupling, 
the D--brane charge, {\it etc.,}) related by conformal invariance.  
We need to find a more general
relationship between these couplings which represents the analogue of
the ``off--critical'' situation. Then we could perform the double
scaling limit in a systematic way.

Such an off--shell approach 
(perhaps {\it e.g.}, ``string field theory'') has not been 
able to help us much in general, but it is
worth noting that we may only need only understand it at tree level in
open string theory to define the off--critical couplings on the
D--brane world volume theory. For open string field theory at tree level, a
fairly complete off--shell definition is understood~\cite{wittenopen},
and this might provide important clues.

A last observation is of course the strength of the effects which are
non--perturbative in the $1/N$ expansion. Ten years ago, we
learned~\cite{shenker} that they were essentially $\e{-1/g_s}$, and
this is one of the features which survives in critical string theory
today. In that context, this form arose as the action of the
``instanton'' representing tunneling of a matrix eigenvalue out of the
potential well containing the ``Dyson gas'' of $N$ eigenvalues. In
critical string theory, we know that the D--branes produce these
effects because their tension (mass per unit volume) goes like
$1/g_s$, and so their appearance in loops will produce such
non--perturbative effects.

Whatever the description of the full $1/N$ expansion of the AdS/CFT
correspondence turns out to be, it is clear that these
non--perturbative contributions will arise again, and this time
clearly in terms of D--branes: Once we allow the full string theory to
come into play in this AdS${\times}{\rm sphere}$ background with
non--trivial R--R charge turned on, the $\e{-1/g_s}$ contributions
of D--branes will naturally appear in virtual loop ``instanton''
corrections due to {\it e.g.,} pair creation in the R--R background field. The
analogy is very complete in this case: in the maximal Abelian subgroup of the
$SU(N)$ gauge theory, the D--branes are in one--to--one correspondence
with the eigenvalues of the matrices.

\section{{\it Fugue:} $SO(32)$ String Duality and the Role of Extended Objects}
Let us discuss one of the string duality pairs discovered in 1995, the system of
the $SO(32)$ Type I/Heterotic strings~\cite{duality}.
\subsection{Dual Actions}
Consider the low energy ($\alpha^\prime{\to}0$) limit, where we study
only the massless spacetime fields produced by the string, 
(the others of course having
infinite mass in this limit). The effective action for the fields in
both cases is a ten dimensional ${\cal N}{=}1$ supersymmetric
gravity theory coupled to Yang Mills.

In the case of the heterotic string, the low energy fields are the
metric $g^{\phantom{2}}_{MN}$, the dilaton $\Phi$, the antisymmetric tensor field
$B_{MN}$, and the gauge field~$A_M$. At tree level in string loop
perturbation theory, the bosonic part of the effective action is (in
the ``string frame''~\cite{heter,chs}):
\begin{equation}
\label{hetsugra}
S_{\rm H}={1\over2\kappa^2_0}\int
d^{10}x\sqrt{-g}e^{-2\Phi}\left(R+4(\nabla\Phi)^2-{1\over12}H^2
-{\alpha^\prime\over4} {\rm Tr}F^2\right),
\end{equation}
where $\kappa^2_0{=}64\pi^7(\alpha^\prime)^4$, and
\begin{equation}
\label{corrections}
H=dB+\alpha^\prime\left(
\omega^L_3(\Omega_-)-{1\over4}\omega_3^{YM}(A)\right)+\cdots
\end{equation}
and $\omega_3$ is the Chern--Simons three form with normalization such that~\cite{anom}
\begin{equation}
\label{bianchi}
dH=\alpha^\prime({\rm Tr}R\wedge R-{1\over4}{\rm Tr}F\wedge F).
\end{equation}
$\Omega_-$ is the usual spin connection modified additively by the
three form $H$ and the dilaton $\Phi$: \be\Omega_{\pm
M}^{ab}=\omega_M^{ab}\pm H_M^{ab}+{1\over2}\left(e^{Na}e^b_M
\partial_N\Phi-e^{Nb}e^a_M\partial_N\Phi\right), \ee where the indices
$(a,b)$ are tangent space indices, to which we refer spacetime indices
using the standard vielbien $e^a_M$.

At this level the duality to the $SO(32)$ type I string is carried out
by the following transformation:
\begin{equation}
\label{dualize}
g^{\phantom{2}}_{MN}\to e^\Phi
g^{\phantom{2}}_{MN},\quad\Phi\to-\Phi,\quad H_{LMN}\to H_{LMN},\quad
A_M\to A_M.
\end{equation}
As the string coupling in each case is $g_s{=}\e{\phi}$, where $\phi$
is the expectation value of the dilaton $\Phi$, this is a duality
relating the strong and weak coupling limits of each theory.  The 
resulting action after transforming the heterotic action~(\ref{hetsugra}) 
with~(\ref{dualize}) is:
\begin{equation}
\label{onesugra}
S_{\rm I}={1\over2\kappa^2_0}\int
d^{10}x\sqrt{-g}\left[e^{-2\Phi}\left(R+4(\nabla\Phi)^2\right)-
{1\over12}H^2-{\alpha^\prime\over4\pi}e^{-\Phi}
{\rm Tr}F^2\right].
\end{equation}
  This is
the type~I string effective low energy tree level action. The fields
$g^{\phantom{2}}_{MN}$ and $\Phi$ are still from the NS--NS sector of the theory,
while $H$ is the field strength of the R--R antisymmetric tensor field
$A_{MN}$, up to important $\alpha^\prime$ corrections which may be
deduced from eqn.(\ref{corrections}).

The Yang--Mills term for the gauge field $A_M$ is multiplied by
$\e{-\Phi}$, instead of $\e{-2\Phi}$ as was the case for the heterotic
action (\ref{hetsugra}). This is a reflection of the fact that the
gauge fields arise at closed string tree level (sphere) in the
heterotic string theory, while they arise at open string tree level
(disk) in type~I string theory. (String perturbative diagrams are
two dimensional world sheets with $h$ handles, $b$ boundaries and $c$
crosscaps, and come with a factor $g_s^{2h-2+b+c}$.)

The $H^2$ term for the R--R sector has no factor of $\e{\Phi}$ at
all. This reflects the fact that this term does not arise in (NS--NS)
closed or open string world sheet perturbation theory, a fact which is
both a blessing (for string duality) and a curse
(for computing perturbatively effectively in
R--R backgrounds). It is crucial to notice that the NS--NS two form
field $B^{(2)}_{\rm H}$ of the heterotic string has transformed into the R--R
two form field $B^{(2)}_{\rm I}$ of the type~I string theory. This means that
objects which charged under these fields are exchanged under duality.

\subsection{The Logic of Duality}
One of the first checks of duality statements (as far as they were
understood) was to examine the behaviour of 
the carriers of the basic degrees of freedom of each theory. 

The reasoning is quite general and is as follows: Imagine that it is
suggested that two theories, theory A and theory B, are dual to each
other. To be precise let us imagine that the duality is organized  by
a single parameter $\lambda$ such that as it gets small, the physics is
best described by A, and if $\lambda$ is large, theory B. Let us start
by considering theory A, and so $\lambda$ is small. Then the defining
carriers of the basic degrees of freedom of theory A should be the
lightest (easiest to excite) states in the theory. The carriers of the
dual degrees of freedom (those of theory B) should be present in
theory A somewhere, but infinitely heavy, and hence not elementary
excitations of the vacuum. Now if the duality is true, then the masses
of both the A carriers and the B carriers should be a function of the
parameter $\lambda$ in such a way that when $\lambda$ is small the A
carriers are light and the B carriers are heavy, and {\it vice--versa}
when $\lambda$ is large.

A duality which was already well known in closed string theory for some
time before strong/weak coupling duality is T--duality~\cite{tdual}. 
In that case, one can compute quite simply in string
perturbation theory that the situation is exactly as we described in
the preceding paragraph. Theory A is a string propagating on a circle
of radius R. The spectrum in one dimension lower has two types of
states whose masses are functions of $R$, the ``momentum states''
whose masses go like $n/R$, and the ``winding states'', whose masses
go like $wR/\alpha^\prime$, where $n,w$ are integers. The former come
from the usual quantization of a particle (string centre of mass) in a
box of size $R$, while the latter come from the fact that the string
can wrap around the box (wind around the circle), because it is an
extended object. For large $R$, the description is best done in terms
of string theory A: the light degrees of freedom are simply the
center of mass degrees of freedom of the string itself, the momentum
modes. For small $R$, those states become heavy, while instead the
winding modes become light. These should now be thought of as the
basic degrees of freedom, realizing a new theory, B. Theory B turns
out not be be a mysterious theory, but simply a closed string
propagating on a circle of radius $R^\prime{=}\alpha^\prime/R$, as can
be seen in the complete spectrum by making the substitution~\cite{tdual,dbranesi,dhs}.

In that example of duality, the parameter $\lambda$ was the inverse
radius of a circle, and the whole discussion can be carried out and
checked in string perturbation theory (weak coupling). What happens if
the conjectured duality is a strong/weak coupling duality? In other
words, what if $\lambda$ is the strength of the coupling of the theory? In
that case, if our techniques for computation are limited to
perturbation theory about $\lambda{=}0$ (which they often are) then we
have a problem. How can we check that the spectrum at strong coupling
is such that the dual degrees of freedom become the lightest? 

In general, with no direct means of computing at strong coupling, we
have no way of actually testing a strong/weak duality statement, as
all of our perturbative tools are useless. However, when we have
extended supersymmetry present, we have an additional lever to
pull. The extended supersymmetry algebra admits central extensions,
which are terms which commute with all of the elements of the
algebra. Such extensions are equal to a charge that objects in the
theory can carry.  As a direct consequence of the algebra, there is a
lower bound on the mass of states in terms of this charge. This is the
BPS bound~\cite{bps,edolive}.  States which are
annihilated by some linear combination of the supersymmetries saturate
this bound and are called ``BPS states''. The formula for the mass and
charge of these states is not subject to quantum corrections
\cite{edolive} and so if
a weak coupling ({\it i.e.,} perturbative) computation of the BPS
spectrum can be carried out, the results may be extended to arbitrary
coupling.

So in our duality discussion, if we have extended supersymmetry, we
have a hope of computing the masses for arbitrary coupling of at least
the BPS part of the spectrum, secure in the knowledge that this part
of the spectrum is true for all values of the coupling.
 Generically, there should be states in the
theory which have mass which behaves inversely with the coupling
($M{\sim}1/\lambda^x, x{>}0$), such that at strong coupling, they
become the light stable states in the theory, defining the ``dual''
degrees of freedom. Such a behaviour for the mass of objects in a
field theory context is found in the spectrum of solitons~\cite{montolive}.

\subsection{Dual Strings}
The technology for carrying out this computation in the string theory
context to test duality centered around a similar
expectation~\cite{atish,hull,harvey,sen}. The
appropriate dual objects will be solitons~\cite{ramzi} of the theory and should be extended solutions in the form of strings, for reasons which will be
clear shortly.

One way to start is to notice that the heterotic supergravity has the
following solution~\cite{fundstring}: 
\bea
\label{fundamental}
ds^2&=&V(r)^{-1}(-dt^2+dx_1^2)+dr^2+r^2d\Omega^2_7 
\\ \e{-2\Phi}&=&1+{M\over r^6}\equiv V(r),\quad\quad
B_{tx^1}=-V(r) 
\eea where $d\Omega_7^2$ is the metric on a unit round $S^7$, with volume
$\omega_7$, and 
\be
\label{mass}
M{=}{N\kappa^2_0\over 3(2\pi\alpha^\prime)\omega_7.}
\ee
Meanwhile the type~I supergravity has the following solution: 
\bea
\label{solitonic}
ds^2&=&V(r)^{-{1\over2}}(-dt^2+dx_1^2)
+V(r)^{1\over2}\left(dr^2+r^2d\Omega^2_7\right)  
\\
\e{2\Phi}&=&1+{M\over r^6}\equiv V(r),\quad\quad
B_{tx^1}=-V(r) 
\eea

In both cases, the solution represents a one dimensional extended
object stretched along the $x_1$ direction. 
Taking into account an
expectation value $\phi$ for the dilaton, a calculation of the tension
(ADM mass per unit length) in each case gives,
for the heterotic supergravity's solution and the type~I
supergravity's solution respectively: 
\be
\label{fundmassi}
T^F_1={N\over2\pi\alpha^\prime}={\mu^F_1\over\sqrt{2}\kappa_0},
\quad\quad T^D_1={N\over 2\pi\alpha^\prime g_s}={\mu^D_1\over\sqrt{2}\kappa_0 g_s},
\ee
in string frame, and
\be
\label{fundmassii}
T^F_1={N\over2\pi\alpha^\prime}g_s^{1\over2},
\quad\quad T^D_1={N\over 2\pi\alpha^\prime}g_s^{-{1\over2}}
\ee 
in Einstein frame. Here, $\mu^{F,D}_1$ are the NS--NS and R--R H--charges, respectively.

\bigskip
\hskip-0.7cm\fbox{\parbox{4.65in}{{\it Obbligato:} 
The ADM mass per unit length is computed in the Einstein frame metric, obtained 
from the string frame metric by multiplying it by $\e{\Phi/2}$. 
An expectation value $\phi$ for the dilaton rescales the constant $\kappa_0$
to $\kappa_0\e{\phi}{=}\kappa_0g_s$.}}
\medskip

The quantities on the left hand side are
the smallest masses allowed by the ten dimensional supersymmetry
algebra, where $Z_1{=}\mu_1/(\sqrt{2}\kappa_0)$ is the central charge. 
Furthermore, the solutions
are annihilated by half of the spacetime supersymmetries~\cite{fundstring} 
(only half
the number of Killing spinors can be defined in the presence of these
solutions) in each case. They are BPS vacuum states.

The interpretation of these solutions is as follows: In the heterotic
supergravity, solution (\ref{fundamental}) represents the fields
around an infinite fundamental heterotic string. The existence of this
solution ensures the self--consistency of the string theory: The
fundamental string generates the 
quanta of the fields $g^{\phantom{2}}_{MN}, \Phi$
and $B_{MN}$, and therefore the effective low energy action for those
fields should admit a solution representing the string. Due to the
mass formulae~(\ref{fundmassi}--\ref{fundmassii}) 
we see that at weak string coupling the
string is light.  It should also be pointed out that the surface
$r{=}0$, representing the core of the string, is actually
singular. This was thought to mean that one simply has to add a delta
function source (the string itself) to the solution in order to make
it a complete solution of the equations of motion, leaving conformal
field theory to supply the missing  description there. The form of the dilaton
shows that the theory is weakly coupled at the core, and so this is consistent. 
(See later.)

In the type~I supergravity, the solution (\ref{solitonic}) represents
a special type of soliton. Its 
(string frame) mass goes like the inverse of a single
power of the string coupling, and not the inverse squared, as is
familiar from solitons in many other contexts.  However, it is
infinitely heavy at weak type~I coupling, and therefore does not
contribute to the perturbative type~I string spectrum.  As
$g_s{\to}\infty$ however, the solution becomes lighter than the type~I
string. This can be seen because the duality transformation (which
exchanges the two solutions) tells us that in the effective
heterotic theory, it is a light object.

This suggests that the solitonic solution is actually the heterotic
string, hiding in the the type~I theory, waiting to come down to zero
mass at infinite coupling and take over the job of dominating the
spectrum. Of course, for consistency, what should actually be checked is that the
putative dual string actually generates the massless spectrum of the
heterotic string as it vibrates. Hence we look for the ``collective
motions'' of the soliton {\it i.e.,} those deformations of that vacuum
solution which have zero cost. We will do that in the next subsection. 

It is worth pausing here to see what other consequences of the duality
map we can explore in the light of this discussion. What we have done
is found a pair of solutions which are mapped to each other under the
duality map, and we have a satisfactory interpretation of the physics
that they represent.

Are there other duality pairs which have interpretations in terms of
duality? The answer is yes. There exist other solutions to the type~I
(and more generally, type II) equations of motion with the interesting
property that they carry R--R charges. Some of them are
generalizations of black holes, and have been called ``$p$--branes''
in general, being $p$--dimensional objects~\cite{pbranes}. 
They couple to R--R forms of
rank $p{+}1$. One of the features of  those solutions is that they
often contain singularities at their core, and are hence generically
 not smooth
soliton solutions.
So their full
interpretation was not fully understood for some time. 

In fact, their
singular nature might have been used as an argument for discarding them as
pathological solutions of the equations of motion. As 
many of these solutions are mapped~\cite{isstrings}  by strong/weak coupling duality 
to smooth NS--NS charged soliton solutions of the heterotic string theory, such a neglect
would be hard to justify: 
One does not throw out smooth soliton solutions, but considers them as
additional sectors of the theory. Therefore duality implies that the
``$p$--brane'' extended solutions are necessarily included in the
type~I and type~II spectra, and must have some interpretation~\cite{isstrings}.
At the time, this duality argument supplied complementary 
 evidence~\cite{stromingerii} that R--R extended solitons were needed to understand 
type~II spectra.

Duality suggests that these
extended objects were an intrinsic part of the theory at least as
important as the other extended solutions, including strings. 
The full spectrum of R--R charged extended objects must 
play an important role because of
their relation to NS--NS objects. 

\subsection{Collective Motions and World Volume Theories}

It is now a familiar story that the collective motions of an extended
vacuum solution localized in some dimensions are described by an
effective theory on its world--volume~\cite{callan}.

One way of seeing this is as follows: By
placing the $p{+}1$ dimensional extended object at a position ${x^m}$
($m{=}p{+}1,\cdots,D$) in the $D{-}p$ dimensional transverse space, one
breaks the translational symmetry of that space. The Lorentz group 
decomposes as: $SO(1,D{-}p)\supset SO(1,p){\times}SO(D{-}p{-}1)$, leaving a Lorentz
symmetry in the space filled by the world--volume of the brane.

\begin{figure}[ht]
\hskip1.0cm
\psfig{figure=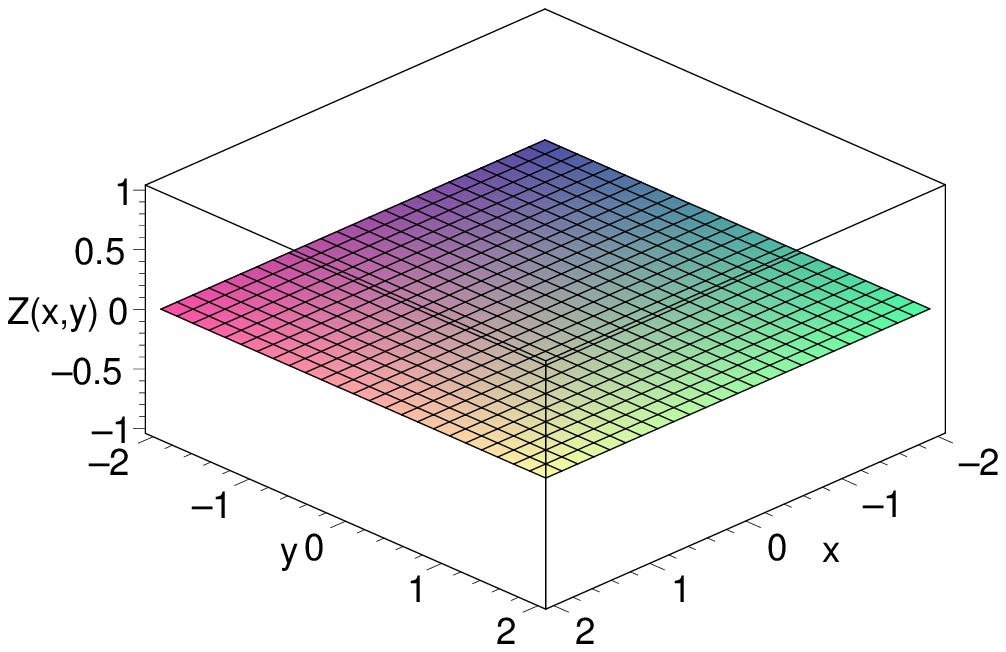,height=1.5in}\psfig{figure=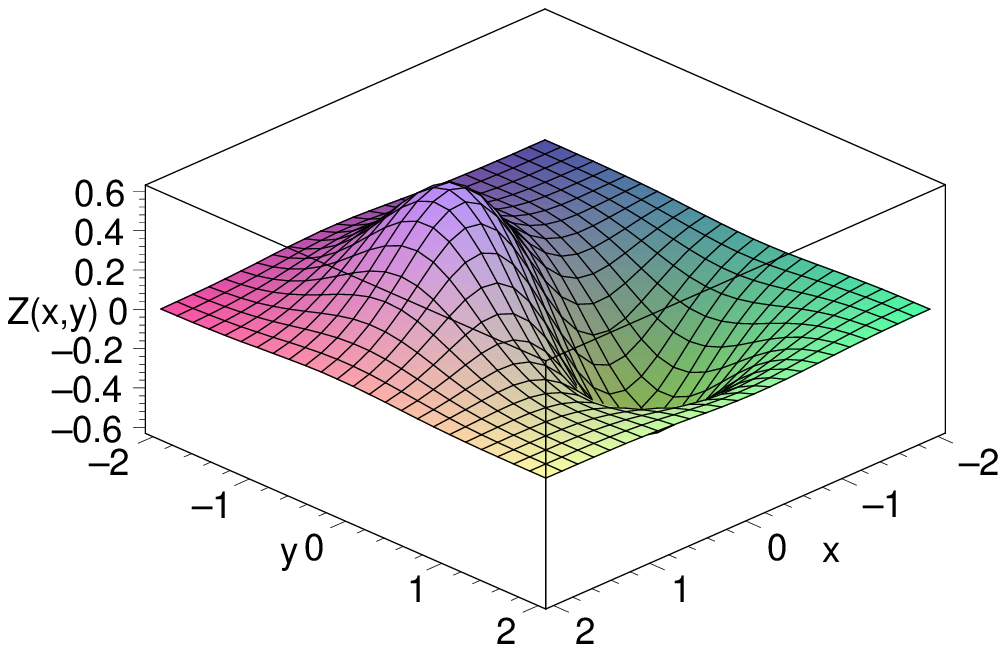,height=1.5in}
\caption{D2--brane: $Z(x,y){=}0$; $\quad$ Notes on D2--brane:
$Z(x,y)=(x{-}y)\e{-(x^2+y^2)}$}
\label{fig:d2brane}
\end{figure}

However, there is still the freedom to redefine the position, $x^m$, by
shifting to another equivalent point. This freedom to
``move'' the object should cost no energy, (it is a trivial redefinition
of the vacuum) and is therefore a simple ``collective motion'' of the
configuration. This is summarized in terms of a collection of $D{-}p$ scalars
$\phi^m(y)$, which are functions of the position, $y^\mu$ in the
remaining $p+1$ dimensional spacetime in which $SO(1,p)$ invariance is
still preserved. In the full theory, these are scalar fields ---Goldstone  bosons--- 
living on the world--volume. Being functions of the position $y^\mu$ means that
we can reconstruct the shape of the object in spacetime, as should
be clear from figure (\ref{fig:d2brane}), where the shape of the
D2--brane shown is described by the field configuration
$Z(x,y){=}(x{-}y)\e{-(x^2+y^2)}$.

The story extends to the case of supersymmetry. The supertranslation
generators are ``broken'', and a family of fermions $\psi(y)$ propagate on
the world volume in exchange for the broken symmetry. The fermions and
bosons form supermultiplets according to the amount of supersymmetry
which was preserved, now reduced to the $p{+}1$ dimensional
spacetime. 

Now, let us be a bit more careful. The supergenerators
in ten dimensional spacetime without the brane present close on
momentum $P^M$. For example, for the case ${\cal N}{=}1$ we have: \be
\label{susie}
\{\Q_\alpha,\Q_\beta\}=({\cal P}\Gamma_M{\cal C})_{\alpha\beta}P^M \ee
Here, $\Gamma_M$ are the gamma matrices satisfying the Clifford
algebra, $\alpha,\beta$ are spinor indices, ${\cal C}$ is the charge
conjugation operator and the projector ${\cal P}\equiv
(1{+}\Gamma_{11})/2$. $\Gamma_{11}\equiv\Gamma_0\Gamma_1\cdots\Gamma_9$.
In the presence of a $p$--brane, the supergenerators no longer close
on the momentum, and have to be modified~\cite{edolive,hlp,town}.
The possible modifications of the algebra are by ``central
terms'' like \be Z^{M_1\cdots M_p}=Q_p\int_{{\cal M}_{p}}
dX^{M_1}\wedge\cdots\wedge dX^{M_p} \ee where ${\cal M}_{p}$ is the
space over which the $p$--brane is extended and $Q_p$ is its charge,
defined by integrating on the asymptotic $S^{D-p-2}$--sphere
surrounding the $p$--brane at infinity: \be Q_p={1\over {\rm
Vol}[S^{D-p-2}]}\int_{S^{D-p-2}}\e{-2\Phi}{}^*H^{p+2}. \ee It is worth
noting that these terms have dimensions of mass, and so cannot be
carried by massless particles in the spectrum.

The modified
algebra in the case of ${\cal N}{=}1$ is: \be
\label{betty}
\{\Q_\alpha,\Q_\beta\}=({\cal P}\Gamma_M{\cal
C})_{\alpha\beta}\left(P^M+Z^M\right) +({\cal P}\Gamma_{M_1\cdots
M_5}{\cal C})_{\alpha\beta}Z_+^{M_1\cdots M_5} \ee and $p$ is {\sl
only} either $1$ or $5$. The subscript $+$ means the restriction to
the self--dual part. The $\Gamma_{M_1\cdots M_p}$ are the
antisymmetrized products of $p$ $\Gamma$--matrices. That (\ref{betty})
is the appropriate form for the ${\cal N}{=}1$ case follows from the
fact that~\cite{townreview} 
the supercharges $\Q_\alpha$ are 16 component spinors, and
so the left hand side of the equation has $16{\times}17/2{=}136$ real
components. The right hand side has $10{+}10!/(5!5!2){=}136$ also,
from $Z^M$ and $Z^{M_1\cdots M_5}_+.$ (Classically we can absorb $Z^M$
into $P^M$, and so do not count their components separately. Quantum
mechanically, however, they are distinct objects, and we must account
for the possibility of objects which carry charges under them
separately, by winding, $Z^M$, or center of mass momentum, $P^M$.)

Consider a static $p$--brane, {\it i.e.,} $P^M{=}0$ for all $M$ except
$M{=}0$. Align the brane along the directions $x^1,\cdots,x^p$, and
$x^0$ is time. Then, as the left hand side of (\ref{betty}) is positive,
we have that \be P^0{\geq}|Z^{01\cdots p}|\quad \Rightarrow \quad
T_p\geq |Q_p|, \ee which is the Bogomol 'nyi--Prasad--Sommerfeld bound~\cite{bps}
on massive states in this supersymmetric theory.  
For branes which saturate the bound, we see
that the brane must be annihilated by supersymmetries built out of 
spinors $\epsilon$ satisfying
\be \Gamma_{01\cdots p}\epsilon=\pm\epsilon. \ee The number of
solutions to this equation is 8,
(the sign is correlated with the orientation of the brane) 
and we see that the BPS state is therefore
annihilated by half of the 16 supersymmetries of the vacuum containing
no branes. The other half do not annihilate the vacuum, but instead
act on the solution, appearing as massless fermions propagating on the
world volume~\cite{hlp}. In this way we get a family of fields which propagate on
the world--volume of the brane.

Now in the case of the heterotic supergravity, the solution
(\ref{fundamental}) is the relevant object which carries the $Z^M$
charge $Q_1$. 
For the type~I supergravity, solution (\ref{solitonic}) is
the appropriate object carrying the charge $Q_1$. 
The Lorentz algebra decomposes into
$SO(1,1){\times}SO(8)$ in the presence of this string.
The ten dimensional 16 component spinor decomposes as
${\bf16}{\to}{\bf8}^s_+{\oplus}{\bf8}^c_-$, where the ${\bf8}^{s,c}$
are the spinor and conjugate spinor representations of $SO(8)$, which
have opposite chirality. The $\pm$ subscript denotes the $SO(1,1)$
chirality. As stated before, half of the supersymmetries (let's say
the ${\bf8}^s$) annihilate the solution, while the others do not. They
instead give right moving zero modes $S^\alpha_-$ on the world
volume. These are the superpartners of the eight bosonic scalars $\phi^i$
propagating on the world volume representing position in the
transverse eight dimensions. The supersymmetry which relates them on
the world--volume is the surviving half generated by the chiral
${\bf8}^c_-$.
 
The action for these modes is the Green--Schwarz free action: 
\be
S=\int d^2\sigma \left[T_1\partial_\mu \phi^i\partial^\mu \phi^i
-iS^\alpha_-\rho^\mu\partial_\mu S^\alpha_-\right] \label{GSheterotic}\ee

That we have found with this simple analysis an oriented
supersymmetric string in ten dimension with $(0,8)$ world--sheet
supersymmetry is already a big clue that this soliton is indeed the
dual heterotic string, given the known short--list of superstrings in
ten dimensions. Indeed, we know that we must have in addition the
equivalent of a family of $32$ left--moving fermions $\lambda$ on the
world--volume as well for consistency and that they must be coupled to
a current--algebra of a 496 dimensional Lie Group.  How can we see
this? Well~\cite{atish,sen}, we wrote a neutral solution (\ref{solitonic}) for the
string, and therefore we may write new solutions equivalent to it
by adding gauge fields which are pure gauge at infinity. As the gauge
group of the low energy supergravity is $SO(32)$, we have a 496
parameter family of equivalent solutions of the equations of
motion. On the world volume, this corresponds to a global $SO(32)$
symmetry of the couplings of the effective theory. That the action for
$\lambda$ is a {\it chiral} $SO(32)$ WZNW model~\cite{wznw} requires a touch
more work. This may be analyzed in some more detail by studying~\cite{atish,harvey}
the possbile fluctuations about the background~(\ref{solitonic}), but we will
postpone a detailed analysis until later.

\subsection{Dual Five--Branes}
Now it is clear that there should also be five dimensional extended object
solutions of the equations of motion for at least two reasons. The
first is from examination of the low energy actions. The three form field strength
$H$ can in principle have magnetic charges as well. Its electric
charge (coming from ``electric type'' non--zero components such as
$H_{rtx^1}$) is found by integrating its Hodge dual on a seven sphere
at infinity. This $S^7$ surrounds the stringy objects (lying along the
$x^1$ direction; $r$ is a radial coordinate in the transverse
directions) which we studied last sub--section.  We can instead have
non--zero components of $H$ (coming from ``magnetic type''
componenets $H_{\theta\phi\psi}$) which give a non--zero answer when we integrate
it on an $S^3$. This $S^3$ (with Euler coordinates $\theta,\phi$ and
$\psi$) would surround a five dimensional extended object in ten
dimensions, a ``five--brane''.

Such a solution was written down in the case of the heterotic string
some time ago~\cite{stromingeri,chs}, and is: \bea
ds^2&=&-dt^2+\sum_{i=1}^{5}dx_i^2+V(r)(dr^2+r^2d\Omega_3^2) 
\nonumber \\
\e{2\Phi}&=&1+N\alpha^\prime {(r^2+2\rho^2)\over(r^2+\rho^2)^2}
+O(\alpha^{\prime2})\equiv V(r),\quad H_{\mu\nu\lambda}=
-\epsilon_{\mu\nu\lambda}^{\phantom{\mu}\phantom{\nu}\phantom{\lambda}\sigma}
\partial_\sigma\Phi 
\nonumber  \\ A_\mu&=&\left({r^2\over r^2+\rho^2}\right)
g^{-1}\partial_\mu g,\quad g={1\over
r}\pmatrix{x_6+ix_7&x_8+ix_9\cr x_8-ix_9&x_6-ix_7}
\label{Ffives}\eea where
$r^2=\sum_{i=6}^9x_i^2$.

One striking thing about this solution is that it is essentially an
instanton (localized in the $x^6{-}x^9$ directions) dressed with some
stringy fields. Indeed, an evaluation of the instanton number, using the
metric in (\ref{Ffives}) gives instanton number $N$.
The instanton has scale size $\rho$, and gives a
non--zero contribution to $dH$ via the equation (\ref{bianchi}).

This is truly a smooth soliton solution of the heterotic string
equations of motion. One may compute the ADM tension to be \be
\label{fivecharge}T^F_5={N\over (2\pi)^5(\alpha^\prime)^3g_s^2}=
{\mu^F_5\over\sqrt{2}\kappa_0g_s^2},\ee where $\mu^F_5$ is the NS--NS H--charge. 
The product
of this charge with that~(\ref{fundmassi}) 
of the fundamental string is $\mu^F_5\mu^F_1{=}2\pi n$, 
($n$ integer)
as required for quantum  consistency~\cite{quantized}. 
This follows for the other 1--5 soliton pair
by duality. The fact that we can get the minimum allowed by setting $N{=}1$ for
both the five and one--brane solutions is crucial.
(Note that as a check that we have correctly modified the conventions of various papers
to match ours, the numbers multiplying the~$1/r^2$ part of the metric
should be:
\be
M={\kappa^2 T_5\over\omega_3}=\alpha^\prime N,
\ee which is what we have. (This is the analogue of equation~(\ref{mass}), 
with $\omega_3{=}2\pi^2$ the volume of a round unit $S^3$.)
The value of the tension given in equation~(\ref{fivecharge}) 
is the minimum allowed by the supergravity algebra for the corresponding
H--charge. The $g_s^{-2}$ behaviour is just the type that we are accustomed 
to for ordinary solitons. An analysis of the available Killing spinors of the solution
shows that it is indeed annihilated by half of the supersymmetries,
verifying that it is a BPS state.

We can find a dual solution of the type~I equtions of motion by
using the duality transformation (\ref{dualize}). It is: \bea
ds^2&=&V(r)^{1\over2}
(-dt^2+\sum_{i=1}^{5}dx_i^2)+V(r)^{3\over2}(dr^2+r^2d\Omega_3^2) 
\nonumber \\
\e{-2\Phi}&=&1+N\alpha^\prime {(r^2+2\rho^2)\over(r^2+\rho^2)^2}
+O(\alpha^{\prime2})\equiv V(r),\quad H_{\mu\nu\lambda}=
-\epsilon_{\mu\nu\lambda}^{\phantom{\mu}\phantom{\nu}\phantom{\lambda}\sigma}
\partial_\sigma\Phi 
\nonumber \\ A_\mu&=&\left({r^2\over r^2+\rho^2}
\right)g^{-1}\partial_\mu g,\quad g={1\over
r}\pmatrix{x_6+ix_7&x_8+ix_9\cr x_8-ix_9&x_6-ix_7}.
\label{Dfives}\eea This
solution is also an instanton of scale size $\rho$. 
A computation of its tension gives:
\be
\label{Dfivecharge}
T_5^D={N\over (2\pi)^5(\alpha^\prime)^3g_s}, \ee and we see that it has the $g_s^{-1}$
behaviour that we saw for the tension of the stringy soliton solution
of the type~I theory.   

\bigskip
\hskip-0.7cm\fbox{\parbox{4.65in}{{\it Obbligato:} It is worth
noting that in order to get the correct value of the dual tensions by
transforming with (\ref{dualize}), we must remember that the tensions
and charges have dimensions of (length)$^{-(p+1)}$, and therefore an
extra factor of $g_s^{p+1\over2}$ must be inserted, given that the
duality transformation involves a rescaling of the metric, which is
used to measure (length)$^2$.}}
\bigskip

This is an example of a phenomenon we anticipated earlier. A
well--behaved smooth NS--NS charged soliton solution of the heterotic
string gets mapped to a singular $p$--brane solution of the type~I
with R--R charges.
As stressed before, this strongly suggests~\cite{isstrings} 
 that these singular R--R $p$--brane solutions are important. 
Now, we know that they are all understood within the framework of 
D--branes, a fact which is consistent with the fact that
their tensions are proportional to $g_s^{-1}$, signaling that they 
have a description in tree level open string theory.

\subsection{More Branes From The Other Extended Algebras}
We may consider~\footnote{The reader ought to consult 
ref.~\cite{townreview} for more on
these types of deductions, and how one may also deduce various 
brane intersections 
from closely related algebras.} 
the extended algebra (\ref{betty}) as descending from
the type~IIB extended ${\cal N}{=}2$ algebra by projecting out the
structures which are odd under~$\Omega$.  \bea 
\{\Q_{i\alpha},\Q_{j\beta}\}&=&\delta_{ij}({\cal P}\Gamma_M{\cal
C})_{\alpha\beta}P^M+ ({\cal P}\Gamma_M{\cal C})_{\alpha\beta}{\tilde
Z}^M_{ij} \nonumber \\
&+&\varepsilon_{ij}({\cal P}\Gamma_{M_1M_2M_3}{\cal
C})_{\alpha\beta}Z^{M_1M_2M_3} +\delta_{ij}({\cal P}\Gamma_{M_1\cdots
M_5}{\cal C})_{\alpha\beta}Z_+^{M_1\cdots M_5} 
\nonumber \\
&+&({\cal
P}\Gamma_{M_1\cdots M_5}{\cal C})_{\alpha\beta}{\tilde
Z}_{ij+}^{M_1\cdots M_5}
\label{deardrie} \eea The two supercharges $\Q_{i\alpha}$,
$(i{=}1,2)$ are of the same chirality: There is therefore an $SO(2)$
action on them which can mix them. The $\tilde Z$ are traceless
symmetric tensors of that $SO(2)$ and are therefore doublets. This
is the full spectrum of charges allowed, given that the left hand side
has $32{\times}33/2{=}528$ components, and the left has
$10{+}2{\times}10{+}120{+}126{+}2{\times}126{=}528$.  The ${\cal
N}{=}1$ algebra (\ref{betty}) descends from this by
$\Omega$--projection because one linear combination of the $\tilde
Z$'s (for $p{=}1,5$) is odd under $\Omega$, as is $Z_{M_1M_2M_3}$ and
the five legged $Z_+$. 

For completeness, we list here also the type~IIA extended
superalgebra. The supercharges here are of opposite chirality, and so
there can be no $SO(2)$ rotating them into each other.  \bea
\{\Q_{\alpha},\Q_{\beta}\}&=&(\Gamma_M)_{\alpha\beta}P^M+
(\Gamma_{11})_{\alpha\beta}Z+(\Gamma_M\Gamma_{11})_{\alpha\beta}Z^M
+(\Gamma_{M_1M_2})_{\alpha\beta}Z^{M_1M_2}  
 \nonumber \\
&+&(\Gamma_{M_1\cdots M_4}\Gamma_{11})_{\alpha\beta}Z^{M_1\cdots M_4}
+(\Gamma_{M_1\cdots M_5})_{\alpha\beta} Z^{M_1\cdots M_5}
\label{maude}
 \eea
and we see that $10{+}1{+}10{+}45{+}210{+}252{=}528$, as it should be.

This algebra descends from the extended ${\cal N}{=}1$ superalgebra in
eleven dimensions:
\be
\{\Q_{\alpha},\Q_{\beta}\}=({\cal P}\Gamma_M{\cal
C})_{\alpha\beta}P^M+ ({\cal P}\Gamma_{M_1M_2}{\cal
C})_{\alpha\beta}Z^{M_1M_2} +({\cal P}\Gamma_{M_1\cdots
M_5}{\cal C})_{\alpha\beta} Z^{M_1\cdots M_5} 
\label{emily}
\ee The terms with
$\Gamma_{11}$ in the algebra (\ref{maude}) descend from terms in
(\ref{emily}) with one more index by winding, while the others descend
directly by dimensional reduction. The $\Gamma$ matrices and
associated projectors in (\ref{emily}) are eleven dimensional
quantities. The supercharges are $32$ component spinors and so the
left hand side still has $528$ components, while the right hand side
has $11{+}55{+}462{=}528$.

We now understand that these algebras inform us about certain branes
which can exist in the various theories. The type~IIB has a doublet of
strings, a doublet of five--branes, and a three--brane. Type~IIA has a
zero--brane, a string, a two--brane, a four--brane and five--brane,
which in turn descend from eleven dimensional momentum, wrapping and
direct reduction of the two--brane, and wrapping and direct reduction
of the five--brane.

\section{{\it Trio:} From $p$--Branes to D$p$--Branes}
\subsection{Trouble at the Core?}
Notice that something interesting happens when the scale size $\rho$ of
the instanton goes to zero~\cite{chs,edcomm}. If this were purely gauge
theory, the instanton
would simply be trivial, but here, there is still a lot of
content. Examining the equations with $\rho{=}0$, we see immediately,
that there is potentially a problem as $r{=}0$. There are at least two
interesting things to say:

\begin{itemize}
\item{The solution appears to diverge as $r{\to}0$, but this is not
the case. In this limit, we can neglect the additive 1 in the
expression for $V(r)$ giving $V(r){\sim}N\alpha^\prime/r^2$ and we can
change coordinates near the core of the configuration to a new radial
coordinate~\cite{chs} $\sigma{=}\sqrt{N\alpha^\prime}
\log(x/\sqrt{N\alpha^\prime})$. This gives \bea\label{hetagain}ds^2&=&
-dt^2+\sum_{i=1}^{5}dy_i^2 +\left(d\sigma^2+N\alpha^\prime
d\Omega_3^2\right)\\ \Phi&=&-{\sigma\over\sqrt{N\alpha^\prime}} + {\rm
constant},\quad H{=}-N\alpha^\prime \epsilon_3, \eea the transverse
part of which has the topology $\IR^1{\times}S^3$, where $\epsilon_3$
is the $S^3$ volume form.  We see that the solution is smooth
everywhere.}
\end{itemize}

\begin{figure}[ht]
\hskip2.0cm\psfig{figure=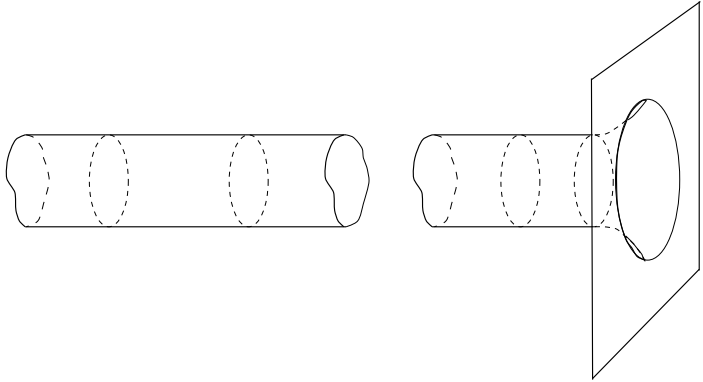,height=2.0in}
\caption{The infinite ``throat'', to the left, is the product
$\IR{\times}S^3$ (see text). To the right is the ``mouth'' region,
connecting the throat onto flat~$\IR^4$.}
\label{fig:throat}
\end{figure}

This product form of the solution is called the ``throat'' of the
solution~\footnote{Actually, a more careful analysis~\cite{mouth} can show the
``mouth'' region where the throat opens up to connect to flat $\IR^4$
as well.}, because if $S^3$ was a circle, the geometry would be that
of a cylinder, as shown in figure \ref{fig:throat}. More generally,
what has happened is that instead of the size of the $S^3$'s of
$\IR^4$ increasing with the radial coordinate, it has frozen to constant
value instead. This frozen value, $R$, is set by $N$ and the string length:
$R^2{=}N\alpha^\prime$. If we keep the product 
$N\alpha^\prime$ large as $\alpha^\prime{\to}0$
and $N{\to}\infty$, we can keep the curvatures all small in the solution,
making sure that at least the curvatures are under control. (Later we will
see that this is apparently not necessary.)

\begin{itemize}
\item{Unfortunately, while the solution is smooth everywhere, we see
that the string coupling $g_s$ diverges as we approach the core,
because $\e{\Phi}$ does. In the new coordinates, this is of course
still true, ($r{\to}0\equiv\sigma{\to}{-}\infty$), 
and we see that the string coupling gets arbitrarily large
as we go down the throat. This ``linear dilaton''~\cite{linear} type behaviour leads
us to wonder whether we have a good description of the solution in
this limit at all (but see later).}
\end{itemize}

Placing the difficulties of interpretation of the strong coupling
behaviour aside for a while, the form of the solution reminds us of an
exact conformal field theory. In this limit, the fact that the
solution has decoupled into such a simple product form can be
exploited. It turns out that there is a
description~\cite{exactcft,chs} (modulo the strong coupling) 
of a string propagating in this target space with that
particular dilaton behaviour. It is the product of an $SU(2)$ WZNW
model~\cite{ff} at level $N$ (the $S^3$) with a Feigin--Fuchs~\cite{ff}
 field of background charge $N$ (for $\sigma$). 
To complete the description, we take the product of this with 
a family of six free fields to represent the flat spacetime
along the brane. This simple description is intriguing, and deserves better 
understanding, especially since the conformal field theory seems to contain
a feature with a problematic interpretation: the string coupling becomes strong.
We will return to this later.

Since these properties were noticed, we have a new handle on the whole
problem: heterotic/type~I duality. The fact that the string coupling
is diverging near the core of our zero--size instanton should lead us
to wonder if there is a better description in terms of the dual
solution in the type~I theory. Examining the form of the dual solution
(\ref{Dfives}), we see that in the limit where $\rho{=}0$:
\begin{itemize}
\item We cannot find a change of variables which will remove the
singularity at $r{=}0$. The solution is singular there.

\item The string coupling gets arbitrarily weak as we approach the
core of the solution.
\end{itemize}

So there is a duality between the good and the bad news in each
case. Here we have a singularity (signaling that the $\alpha^\prime{=}0$ limit 
has failed to capture some of the physics), but the string loop expansion in 
$g_s$ is under control, while on the heterotic side, the solution is smooth, and hence the
$\alpha^\prime$ expansion is under control, but we have no control
over the string coupling.

Is this progress? At $\rho{=}0$, both sides of the discussion are
apparently deficient in some way. It seems that there is still a component of the
discussion missing. 

\subsection{Clues From Anomalies}
Let us try to anticipate how  new physics might arise to supply the
missing sector. An important constraint on the allowed sectors in the
theory is supplied by the anomaly. In the effective six dimensional
theory on the world volume of the five--branes, we have ${\cal N}{=}1$
supersymmetry. To constrain our dynamics on the world--volume with the
anomaly properly, we should actually make the transverse space into a
compact manifold which preserves the amount of supersymmetry which we
require. The only manifold with this property is called $K3$. It is a
``Calabi--Yau'' manifold, which for our purposes is simply a manifold
with a K\"ahler structure (which is an extra--special form of a complex
structure) with $SU(2)$ holonomy. This merely means that it preserves
half of the spacetime superymmetries~\cite{aspinwall}.

The allowed multiplets which can appear in ${\cal N}{=}1$ $D{=}6$ are
called ``vector multiplets'', ``hypermulitplets'' and ``tensor
multiplets''. The bosonic parts of these multiplets all have four
field theoretic degrees of freedom. Their transformation properties
under the $SU(2){\times}SU(2)$ covering group of the $SO(4)$ little
group of $SO(5,1)$ are $\bf(2,2)$, $4{\bf(1,1)}$ and ${\bf(3,1)}{+}{\bf(1,1)}$
respectively. So the hypermultiplet is simply four scalars, while the
tensor is an antiself--dual antisymmetric field $B_{\mu\nu}^-$ plus a
scalar.

The gravity supermultiplet has bosonic part ${\bf(3,3)}{+}{\bf(1,1)}$,
the graviton $G_{\mu\nu}$ and dilaton $\Phi$, and is completed
by a self--dual antisymmetric tensor $B_{\mu\nu}^+$ $\bf(1,3)$, and
three other scalars to contain the dilaton in a hypermultiplet. The
familiar $B_{\mu\nu}$ field is assembled from the $B_{\mu\nu}^+$ $\bf(1,3)$ and
a $B_{\mu\nu}^-$ $\bf(3,1)$ from an ordinary tensor multiplet. We
will therefore take it as given that this tensor multiplet is always
in the theory. Any tensor multiplets mentioned in a spectrum hereafter
are understood to be in addition to this one.
 
Gauge and gravitational 
anomalies~\cite{anomalies} constrain the allowed content of the theory. Let us denote
the number of vectors, hypers and (extra) tensors by $n_V$, $n_H$ and
$n_T$, respectively. A necessary (but far from sufficient!)
condition that the anomalies vanish  is
that \be n_H-n_V=244-29n_T
\label{anomaly}
\ee (this is actually the coefficient of the irreducible ${\rm Tr}R^4$
term in the gravitational anomaly.)
 
How might we make this work for us? Well, away from the small
instanton limit, we may construct (for example) a consistent heterotic
or type~I vacuum using the fivebrane as follows: They are instantons
and hence sources of $F{\wedge}F$. We saw from the supergravity
(equations (\ref{hetsugra}) and (\ref{onesugra}) with (\ref{bianchi}))
that this means that they are magnetic sources for $H$--charge (NS--NS
sector in heterotic and R--R in type~I) supported in $x^0{-}x^5$. Now that we
have a compact transverse space $x^6{-}x^9$, the field equations for
$H$ become a Gauss--Law type condition on its field lines, requiring
its sources to be accompanied by sinks within the compact
spacetime. Fortuitously, the same equation (\ref{bianchi}) which told
us that instantons are a source also tells us that non--trivial
$R{\wedge}R$ is a sink. $K3$ has 24 units of this
(its Euler number) and so we can satisfy the equations of motion by
having 24 fivebranes present at arbitrary positions in the $K3$. As
$SO(32)$ instantons, they break the gauge group. We can choose how
they are embedded into $SO(32)$ (choose the ``gauge bundle'') in many
ways. The minimal way is to embed them all into the same $SU(2)$
subgroup, breaking $SO(32){\supset}SO(28){\times}SO(4)$ to
$SO(28){\times}SU(2)$. An index theorem~\cite{index} tells us how many ways there
are of doing this, which is~615. (This number includes the 24
positions of the instantons in $K3$ and their $SU(2)$ orientations.)
In the six dimensional model, this translates into a number of extra
hypermultiplets which parameterize these distinct choices. Their
transformation properties under $SO(28){\times}SU(2)$ are $10{\bf
(28,2)}{+}45{\bf(1,1)}$. It is important to note that $K3$ also comes
with 80 numbers (``moduli'') which specify its shape, which translates
into 20 hypermultiplets. (The fact that they are naturally in groups
of four is a consequence of the hyperK\"ahler structure of~$K3$.)

So we have $n_V{=}{\rm dim}[SO(28)]{+}{\rm dim}[SU(2)]{=}381$, $n_T$=0
and $n_H{=}625$, which satisfies (\ref{anomaly}). Of course, we should
also check that the other anomaly polynomials vanish ---and they do---
but we will not do that here~\cite{erler}.

It should be immediately apparent from the form of equation
(\ref{anomaly}), and the fact that vectors and hypers have the same
number of bosonic components, that starting from a theory with some
content allowed by the anomaly, there is the possibility to move to a
new theory (or, more properly, a new {\it branch} of the theory) where
we have either
\begin{itemize}
\item increased the
number of vectors by the same amount by which we have decreased the
number of hypers (or {\it vice--versa}), or similarly 

\item{exchanged some number, $n$, of hypers with 29$n$ tensors.}

\end{itemize}
Both mechanisms, especially the first, should remind us of the Higgs
mechanism, and that is the key.

\subsection{An Economical Resolution}
As the anomaly is a constraint on the {\it full} quantum theory and
not just our classical analysis, whatever new physics might occur
should give a spectrum consistent with the anomaly. So we can conclude
the following: In either the heterotic or the type~I picture, as we
approach the limit of small instantons, we seem to get a singularity
in the supergravity description. We must recall, however, that the
supergravity description is an {\it effective} description of the massless
degrees of freedom of the theory, and we have implicitly integrated out all of
the massive degrees of freedom. 

We therefore must consider the
possibility~\cite{stromingerii} that the singularity we are encountering is {\it simply a result
of having unintentionally integrated out fields which are becoming massless in the
small instanton limit.} Put another way, the scale size of the
instanton might correspond to the vacuum expectation value 
(``vev'') of a scalar in a hyper
mulitplet, or the mass of a vector or tensor multiplet. The anomaly
(equation (\ref{anomaly}) and the other polynomials) allows a Higgs
mechanism to take place and permit such new massless fields to appear.

We shall see that this is precisely what happens~\cite{edsmall}. 
The small instantons are D5--branes, for which an enhanced gauge 
symmetry lives naturally on their world--volume, carried by vectors. 
D5--branes are the description of the small $SO(32)$ instantons we
seek~\footnote{Actually, the possibility of extra {\it tensors}
appearing to resolve the singlurity also occurs, but for the small
$E_8{\times}E_8$ instantons~\cite{origanor}. 
That interesting story will have to wait
for another time, due to lack of space.}. 

So we see that we can complete the description of the fivebranes by adding
a gauge theory when the type~I supergravity description loses its
predictability. On the other hand, the heterotic description 
apparently failed us
in a different way: we simply did not know how to make an honest
interpretation of the infinitely long throat because the linear
dilaton placed us at arbitrarily strong coupling down there (see
later, however).

This is an example of an effect which can take place in the heterotic
string theory for any value of the coupling. Traditional perturbative
heterotic methods cannot see this new massless sector at all. Indeed,
this forces us to re--examine many of the conclusions made about the
particle physics phenomenology of the heterotic string, as not only do
new gauge groups appear, but new types of matter representations
also. This is both a blessing and a curse for phenomenology as we
understand it, as explained in the lectures of J.~Louis in this
school. (See also the lectures of B. Greene on F--theory, where the possible 
non--perturbative gauge groups and matter representations due to small 
heterotic instantons are described in a larger framework.)

\section{Type~I String Theory Under the Microscope~I: Dual Strings}
\subsection{D9--Branes}
In modern parlance~\cite{gojoe}, we look at the type~I string action
(\ref{onesugra}) as follows: We started with the type~IIB string and
did a very simple orientifold, dividing by the $\IZ_2$ action of world
sheet parity $\Omega$. As $\Omega$ acts on the type~IIB string
everywhere in spacetime we say that we have an ``O9--plane'' filling
space, which has $-16$ units of charge of the R--R ten form potential,
$A^{(10)}$. The form has no contribution to the action
(\ref{onesugra}) in terms of its field strength. Instead, the ``Gauss'
law'' equation of motion simply requires us to cancel this charge by
introducing 16 pairs of D9--branes (open string sectors with
$32{\times}32$ Chan--Paton matrices), which each have $+$1 unit of
charge.

The first three terms in the action reflect the equation of motion for
the type~IIB fields which survived the projection.
The last term is the leading term in the expansion in
$\alpha^\prime$ of the world volume action of the D9--branes. The full
action is the Born--Infeld action, and one can therefore read off the
value of the D9--brane tension as
$\tau_9{=}(2\pi)^{-9}(\alpha^\prime)^{-5}g_s^{-1}$. The gauge symmetry
is $SO(32)$ because 16 pairs of D9--branes give such a gauge symmetry,
because their $U(32)$ gauge symmetry gets projected to $SO(32)$ by the
action of~$\Omega$.

The D9--branes are introduced into the string theory as Neumann
boundary conditions on the strings in all spatial directions
$x^1{-}x^9$. The world--volumes of the D9--branes fill the whole of
spacetime.  On the world--volume lives a gauge theory $SO(32)$.

\subsection{D9--Branes and D1--Branes: The Dual Heterotic String}
\begin{figure}[ht]
\hskip2.0cm
\psfig{figure=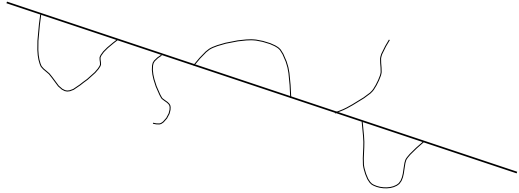,height=1.0in}
\caption{The type~I D1--Brane and some of the strings that define it.}
\label{Dstring}
\end{figure}

A D1--brane is placed into the theory extended along the $x^1$
direction by adding Dirichlet boundary conditions. We can ask for
strings to end at a point in the $x^2{-}x^9$ directions, leaving then
free (Neumann) in the $x^1$ direction. The collective coordinates of
the soliton thus described are in the zero modes of the strings
connecting the D1--brane to itself (``1--1 strings''), and those
connecting it to the D9--branes (``1--9 strings''), as shown in 
figure~\ref{Dstring}.  These zero modes are fields propagating on the
(1{+}1)--dimensional world--volume of the D1--brane.

In the presence of the brane, the spacetime Lorentz group breaks as
\be SO(1,9){\to}SO(1,1){\times}SO(8), \ee
where the $SO(1,1)$ refers
to the world--volume of the D1--brane. The spacetime supercharges
decompose as ${\bf16}{\to}{\bf8}^s_+{\oplus}{\bf8}^c_-$, where the
$8^{s,c}$ are the spinor and conjugate spinor representations of the
$SO(8)$ and the subscripts denote $SO(1,1)$ charge (chirality). The
D1--brane is annihilated by one of these (choose the first), and the
other remains as a supercharge of the (0,8) supersymmetric model in
1+1 dimensions representing the collective fluctuations of the brane.

Let us be fairly general~\cite{edjoe,ulf,banksseib}, and add $N$ 
D1--branes to the theory at one
time. The $1{-}1$ strings break up into two classes. Those with
components in the directions transverse to the D1--branes, and those
with parallel components.  The massless excitations form vectors and
scalars in 2D, and are created as follows.  The latter class
represents the collective motions parallel to the brane ane are vectors
of the $SO(1,1)$ Lorentz symmetry on the brane. As there are 8
Dirichlet--Dirichlet (DD) directions, the Neveu--Schwarz (NS) sector
has zero point energy $-{8\over24}{-}{8\over48}{=}{-}{1\over2}$. The
vectors $A^\mu(x^0,x^1),$ $(\mu{=}0,1)$ come from the excitations of
the Neumann--Neumann (NN) directions: \be
\lambda_V\psi^\mu_{-{1\over2}} |0>\quad{\rm with}\quad \lambda_V=
-\gamma_\Omega^{\phantom{-}}\lambda^T_V\gamma_\Omega^{-1},\quad\mu=0,1.
\label{oneonenn}
\ee Here, $\lambda_V$ is an $N{\times}N$ Chan--Paton matrix, which
must satisfy the conditions shown. The $\gamma_{\Omega}^{\phantom{-}}$
matrices are $N{\times}N$ matrices chosen to represent the action of
$\Omega$ on $\lambda_V$. As a result the vectors carry an $O(N)$ gauge
symmetry.

The transverse fluctuations are a family of eight scalars
$\phi^i(x^0,x^1)$, $(i{=}2,\ldots,8)$ on the world volume. These come from
the 8 DD directions \be \lambda_\phi\psi^i_{-{1\over2}} |0> \quad{\rm
with}\quad \lambda_\phi\to
\gamma_\Omega^{\phantom{-}}\lambda^T_\phi\gamma_\Omega^{-1},\quad
i=2,\ldots,8.
\label{oneonedd}
\ee  The scalars therefore transform in the $N(N{+}1)/2$ dimensional
symmetric tensor representation of the gauge group. 
The $SO(8)$ global symmetry which rotates them into one
another is a simple consequence of the symmetry the brane
configuration and corresponds to the R--symmetry of the chiral ${\cal
N}{=}8$ model we are studying.

The fermionic states $\xi$ from the Ramond (R) sector (with zero point
energy 0, by definition) are built on the vacua formed by the zero
modes $\psi_0^i,\quad i{=}0,\ldots,9$. This gives the initial $\bf 16$
supercharges mentioned earlier. The GSO projection acts on the vacuum
in this sector as: \be (-1)^F=\Gamma^0\Gamma^1\ldots\Gamma^9,
\label{gso}
\ee
while as $\Omega$ acts as $-1$ on NN strings ({\it i.e.,} in the $(x^0,
x^1)$ directions), it is:
\be
\Omega=\Gamma^2\ldots\Gamma^9.
\label{omegga}\ee
So we have $(-1)^F\xi{=}\xi$ from the GSO projection, and with
$\Omega$, it simply correlates world--sheet chirality with spacetime
chirality: $\Gamma^0\Gamma^1\xi_{\pm}=\pm\xi_\pm$, where $\xi_-$ is in
the ${\bf 8}_c$ of $SO(8)$ and~$\xi_+$ is in the~${\bf 8}_s$. 
They are the superpartners
of $\phi^i$ and $A^\mu$, respectively, carrying
the same $O(N)$ charges as their bosonic superpartners,  ensuring that gauge
symmetry respects supersymmetry. 

The 1--9 strings will also form a family of fields on the world--volume.
There are 8 Dirichlet--Neumann (DN) coordinates, giving ground state
energy~${1\over2}$, and so there are no massless states arising in the
NS sector. The R sector excitations come from the NN $(x^0,x^1)$
system giving just two ground states. In this sector, the GSO
projection is simply $(-1)^F{=}\Gamma^0\Gamma^1$, which picks the
left--moving field~\cite{edjoe}.  As we have gauge group $SO(32)$ from
the D9--branes, we have left--moving fermions $\lambda_+^M$ in the
$\bf{(N,32)}$, where $M$ is an $SO(32)$ (D9--brane) index.

Consider the case of one D1--brane for a moment. 
Then we have no gauge group on the world--volume, as
the vectors are projected out by $\Omega$, and the remaining fields are
simply the eight scalars $\phi^i$, their right moving superpartners
$\xi_-$ in the ${\bf8}_c$ and the 32 left moving fermions
$\lambda_+^M$. This is simply the content of the $SO(32)$ heterotic
string in static gauge where the $\lambda_+^M$ are the current algebra
fermions.  The action for this theory is simply the light--cone gauge
Green--Schwarz action (\ref{GSheterotic}) for the heterotic string
with a current algebra term added.

In the case of the multiple D1--branes, we have a non-abelian
generalisation of that model: \bea S=\int d^2\sigma \,{\rm
Tr}\biggl[T_2{\partial}_\mu \phi^i{\partial}^\mu \phi^i
-iS^T_\alpha{\slash\hskip-0.25cm{\cal D}} S^\alpha-i\lambda_+^M{\cal
D}_-\lambda_+^M+{1\over g^2} F_{\mu\nu}F^{\mu\nu}+\hbox{\rm
extra}\biggr]
\label{MatrixGS}
\eea Here, $g^2{\sim}g_s/\alpha^\prime $ is the effective gauge
coupling of the $(1+1)$--dimensional theory, and
$S^T_\alpha{\equiv}(\xi^\alpha_-,\xi^\alpha_+/g)$, with
${\slash\hskip-0.25cm{\cal D}}{\equiv}({\cal D}_+,{\cal D}_-)$.

In models such as this, 
the ``extra'' terms may be written as a combination of commutators
between the various fields, their precise form determined by gauge
invariance and supersymmetry.  In this example, one such term is
$g^2[\phi^i,\phi^j]^2$ and a similar term for the
$S_\alpha$, and a Yukawa term coupling the $\xi_\pm$ and $\phi^i$.
Such terms constitute the ``scalar potential'' of the
model.

\bigskip
\hskip-0.7cm\fbox{\parbox{4.65in}{{\it Obbligato:} The supersymmetric
vacua of such a gauge theory are those for which the ``scalar potential'' is
identically zero. We can immediately study the classical solutions of
this condition by just treating the vanishing of those terms as an
algebra problem. The space of {\it gauge inequivalent} solutions of
this condition is grandly termed the ``moduli space of classical
vacua''. In general, quantum corrections can modify our answer, but with
the right type or amount of supersymmetry (for example), the classical analysis is
equivalent to the quantum analysis. This moduli space is the space of
allowed values that the fields can take. Given that we have already
realized that the fields on the world--volume of the branes are in
one--to--one correspondence with the geometry that the brane
encounters ---both the embedding space and the shape that it can take
in that embedding space--- evaluating the moduli space of vacua is
equivalent to discovering this geometry. This is the key to many relationships
between geometry and field theory.}}
\bigskip

Turning to the moduli
space of vacua of this model,
we see that the point with gauge symmetry $O(N)$ is a special point of enhanced
gauge symmetry. All of the scalar fields have zero vevs, and so the
commutators (and hence the scalar potential) vanishes
identically. This corresponds to all of the D1--branes being at the
same point, which we have taken to be the origin of the
eight--dimensional space $\IR^8$ parameterized by $\phi^i$,
$i{=}2,\ldots,9$. Generically, the fields can have non--zero vevs, but
we wish to still consider supersymmetric solutions, which is to say we
want the potential still to vanish. A solution is to make
the $\phi^i$'s non--zero, but all commute. In this way, we break the
gauge symmetry down to the ``maximal torus'' (the largest Abelian
subgroup) of $SO(2N)$, which is $U(1)^{N\over2}$. This corresponds to
separating out $N$ D1--branes pairs. This situation is further reducible
however (in contrast to a similar situation for D5--branes~\cite{edsmall,gp}), 
and we may split the D1--brane pairs. The resulting gauge group is $\{0\}$ as
we have seen, and there is one eight component scalar $\phi^i$ left for each of the $N$
D1--branes, representing their transverse positions. (Indeed, that we can Higgs the gauge
group away leaving $N$ scalars follows from the fact that the difference between
the dimensions $N(N{-}1)/2$ of the adjoint and the $N(N{+}1)/2$ of the symmetric is $N$.)

As they all commute, we may find a basis where we simultaneously
diagonalize the $\phi^i$ matrices, putting their eigenvalues down the
diagonal: each eigenvalue represents an individual D1--brane. Notice
that the Weyl group is still a gauge symmetry here, acting to permute
the eigenvalues. This translates into the fact that the theory does not care if we
rearrange the $N$ D1--branes, as they are identical. Therefore the
classical moduli space of vacua is not $(\IR^8)^N$, but
$(\IR^8)^N/S_N$. The action of this $S_N$ will have important
consequences shortly. Notice that we can get special points of $O(n)$
enhanced gauge symmetry whenever $n$ D1--branes coincide, which
corresponds to having $n$ simultaneous eigenvalues in the eight
$\phi^i$'s.

We have not quite finished the job yet, as we have not discussed the
allowed vacua of the superpartners $\xi$ at all. However, this is not
necessary, as we have sought supersymmetric solutions here. Therefore,
their allowed values are determined by the unbroken supersymmetries.
There remains to be determined the allowed values of the
left--moving current algebra fermions $\lambda_+^M$. Up to subtleties
we will mention later, this is simply parameterized by the fact that
they are fundamentals of the D9--brane gauge group $SO(32)$, and hence
parameterize the vector space $V_{32}{\approx}\IR^{32}$ that it acts
on.  So the full moduli space is schematically
$(\IR^8{\times}V_{32})^N/S_N.$

\subsection{From D1--branes to Fundamental Strings}

So we understand now that the type~I supergravity model of the
solitonic heterotic string that we were studying previously represents
the fields around $N$ coincident D1--branes. The world--volume theory of that
soliton has been found more precisely to be our 1+1 dimensional gauge
theory.

Notice that the coupling of the gauge theory is a function of the
type~I string coupling. We had promised that as $g_s$ goes to
infinity, we would arrive at the heterotic string theory. What does
this mean for the 1+1 dimensional model? The 1+1 dimensional coupling
gets strong too, and so as a 1+1 dimensional gauge field theory, it
should flow to the infra--red, presumably to a fixed point.
In the special case of one D1--brane, the conformal field theory that we flow
to is clearly the $(0,8)$ supersymmetric $(c_L,c_R)=(24,12)$ conformal
field theory of the free $SO(32)$ heterotic string, but what of other $N$?

For general $N$, the model is a non--Abelian gauge theory, and
therefore there are potential terms like $g^2[\phi^i,\phi^j]^2$. As
the string coupling goes large, so does~$g$, and this term becomes
very important. Indeed, at infinite $g$, the only way to find
supersymmetric vacua is to force this term to zero by demanding that
the $\phi^i$ all commute, generically (to set them all to zero is
highly non--generic). So in effect, strong coupling forces us out onto
the Abelian (Coulomb) branch again, and the allowed values of the $\phi^i$'s are
in $(\IR^8)^N/S_N$. What is the interpretation of this?

Given that we identify configurations related by the action of $S_N$,
the permutation of the eigenvalues, it is useful to think of
$S_N$ as a sort of discrete gauge symmetry. The usefulness of this
comes when we recall that our world--volume theory arose in
representing the dynamics of a stable closed string made by winding it
about a circle ($x^1$) with a very large radius $R^1$. We have not
discussed that feature much so far, but it is crucial. Indeed, this
model of $N$ D1--branes is indistinguishable from a model of one
D1--brane wound $N$ times around the large circle, or any number of
D1--branes with individual windings distributed among them to make
total winding $N$. The moduli space $(\IR^8)^N/S_N$ encodes precisely
that. The interpretation as $N$ D--branes that can be permuted by
$S_N$ is therefore a small part of the story. The $\phi^i$ are {\sl
matrix--valued} functions $\phi^i(x^1)$ which can go around the circle
(with coordinate $x^1$, recall) and return to their original value
{\it up to an action of an $S_N$ permutation of their eigenvalues}.

\begin{figure}[ht]
\hskip3.0cm\psfig{figure=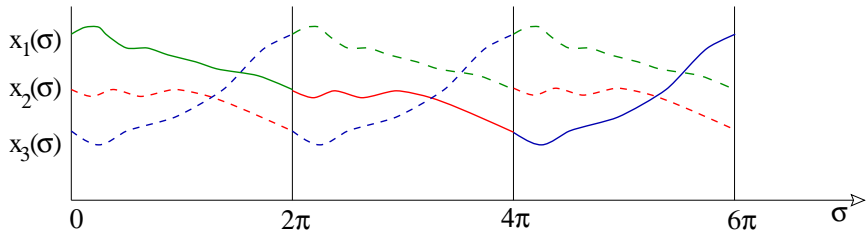,height=1.2in}
\caption{A winding D--string of length three, made by acting with an
$S_3$ twist. The physical string is the solid line, made by gluing the
three functions together. (Green, then red, then blue, for viewers in
colour.) At strong coupling, this string becomes the fundamental
heterotic string with three units of momentum in the discrete
light--cone gauge. (See text.)}
\label{orient}
\end{figure}

Imagine a particular field configuration $\phi^i_n(x)$ with such a
``twisting'' of its boundary conditions by $S_n{\subset}S_N$:
$\phi^i_n(x^1){=}S_n\cdot\phi^i_n(x^1{+}2\pi R^1)$. If the permutation
$S_n$ is irreducible, then we can return to the original matrix
$\phi^i_n(x^1)$ only by going around the circle $n$ times. In other
words, this multi--valued configuration may be written as a
single--valued one by using a circle of radius $n$ times that of the
basic circle: $\phi^i_n(x^1){=}\phi^i_n(x^1{+}2n\pi R^1)$. We have just
described a configuration representing the winding of a single
D1--string $n$ times around the circle. Figure~\ref{orient} shows
the situation for $n{=}3$, covering the circle three times. We have
focussed on a $3{\times}3$ submatrix of $\phi^i$, with entries
$x_1(\sigma),x_2(\sigma)$ and $x_3(\sigma)$, functions which get
permuted every time we go around the circle. Here, $\sigma$ is the
world--sheet spatial coordinate of the string which gets identified
with the circle $x^1$ in this ``static gauge''. The physical string
is the solid line, made by gluing the three functions together.

The beauty of this description is that it is a very economical way of
describing strings of arbitrary shape~\footnote{Recall from section~2.4 that
the shape of the brane in encoded in the actual functional form of the eigenvalues
of the matrix $\phi^i(x^1)$.} in the embedding space $\IR^8$
(wound along a direction $x^1$, with any
amount of winding up to a total of $N$. The shortest length of string
is $2\pi R_1$ and the action of the discrete gauge group $S_N$ glues
together these short strings to make longer strings. We have another
language for this: This (1+1)--dimensional field theory {\it is
actually a light--cone  string field theory}, as it has fields
$\phi(x)$, which create and destroy strings of arbitrary shape in the
transverse $\IR^8$. Notice that in this light--cone interpretation,
the static gauge string with winding $N$ is exchanged for a
light--cone gauge string with momentum $N/R$, as is consistent with
T--duality.

Now all of this could have been said before about the Coulomb branch,
without recourse to the large $g$ limit, but what makes this very
relevant is the fact that at strong coupling, each of the short
strings, which has tension~${\sim}1/g$, becomes a light heterotic
string. Furthermore, as $N$ becomes large it can be shown~\cite{verlindematch}
 that the long strings made by gluing the short strings together have the same
world--volume dynamics as the short strings, and that the interactions
between the strings are generated by the field theory
interactions. Unfortunately, lack of space does not permit us to
describe the full story here, but luckily H. Verlinde will describe
this subject of ``Matrix String Theory'' in his lectures~\cite{microverlinde}. 

Notice that in taking $N$ large, but staying in the
$\alpha^\prime{\to}0$ limit, we have taken ourselves back to the realm
of validity of the original supergravity discussion of the previous
discussion. The description of the $N$ D--brane soliton conglomerate
is well approximated by the supergravity solution
(\ref{solitonic}). However, we have taken the string coupling to be
very large, and therefore (because of the dilaton's behaviour) we are
studying the physics closer and closer to the core of the solution. In
infinite coupling, we should simply exchange this solution for the
heterotic one given in equation (\ref{fundamental}), arriving at the
nieghbourhood of the core of that solution.

As we noticed before, however, the fundamental string core is
singular as a supergravity description. We were supposed to think of
this as a result of the natural breakdown of the description of the
self--consistent solution at weak coupling, and we can replace this
singular core with a delta--function source, leaving conformal field theory
vertex operators to take over the description~\cite{fundstring}. 
Now we see that we do
have a complete description of the missing physics in the $SO(32)$
heterotic supergravity description, it is the
$(\IR^8{\times}V_{32})^N/S_N$ orbifold conformal field theory~\cite{super,juan}.

There is an important subtlety here.  We have seen that the orbifold
conformal field theory describes heterotic strings, but we have not
been careful to check which heterotic string. Recall that they are
indistinguishable at tree level, and therefore the naive large $N$ and
$g$ analysis above is good for either string. Similarly, the
``duality'' to the supergravity solution can be to either heterotic
supergravity fundamental string solution if we are not careful. It
turns out that a more careful treatment of the model~\cite{matrixhet} 
for arbitrary~$g$ and 
radius~$R$
requires that a Wilson line be turned on in the $x^1$ direction
breaking the $SO(32)$ gauge symmetry to $G{=}SO(16){\times}SO(16)$ and
thus dividing the current--algebra fermions $\lambda_+^M$ into two
classes: periodic or antiperiodic as they wind around the circle
$x^1$. The full moduli space of the model is then
$(\IR^8{\times}(V_{16}{\oplus}V_{16}))^N/(S_N{\times}\IZ_2)$, where the
$\IZ_2$ exchanges fermions in either $V_{16}$ factor. In the large
$N,R$ and $g$ limit, it turns out that the long heterotic strings
which are recovered are $E_8{\times}E_8$ heterotic strings! 
 This turns out to be consistent with the fact that the model can be
obtained by compactifying Matrix theory on a line interval: Matrix
theory is the light--cone representation of M--theory and M--theory on
a line interval gives the $E_8{\times}E_8$ heterotic 
string~\footnote{The line interval is explicit in the description if one 
T--dualizes on $x^1$
and works in the dual theory. The action of T--duality combined with
the orientifoldng~$\Omega$ makes the dual circle into a line interval by 
orbifolding. There is an O8--plane at each end of the interval. 
The D9--branes become point--like along the dual circle and turn into D8--branes. This is
called ``type~IA'' string theory, to contrast it with the ordinary type~I theory, which might
be called the ``type~IB'' theory. (The terminology 
is good, as they each decend from each of the 
similarly named type~II theories by simple orientifolds.)
 Local considerations~\cite{edjoe,petred} require that
 there be 8 D8--branes
at each end of the circle, breaking $SO(32)$ to $SO(16){\times}SO(16)$.}

To get the $SO(32)$ heterotic string in this way is more subtle. It
has been shown~\cite{super} that the tree level description of the heterotic string
is indeed identical to the above (including the ``dual'' supergravity
description in terms of the core of a fundamental string), but the
situation is different when the string coupling is turned on: The interacting 
physics is described by  an exotic 2+1 dimensional model. This turns out to
be consistent with the fact that the $SO(32)$ heterotic string arises
as M--theory compactified on a cylinder.

\subsection{Cadenza: Where Is The Fundamental Type~I String?}
Going back to the discussion of the previous section, it is reasonable
to ask after the whereabouts of the solution representing the
fundamental type~I string. Surely, for consistency, there should also
be a solution of the type~I supergravity (\ref{onesugra}) representing the
fields created by the string itself?

It is important to note that when we did this for the heterotic
string, we constructed a stable solution. It was a BPS state, stable
because it represented a closed string wrapped on a circle of large 
 radius. As it is a closed string, it cannot shrink away
to minimize its energy. By contrast, a type~I string cannot be made
stable by the same procedure because it can reduce its energy by
snapping, (still ending on space--filling D9--branes) making shorter
strings in this way. So we cannot make stable strings in this way, and
therefore cannot expect to find a BPS state in the spectrum
corresponding to such a configuration. One does not expect to find it
in the type~IB supergravity as a fundamental string solution, and not
in the heterotic supergravity as a stable soliton solution (although
one might be able to make metastable solutions).

One should not be discouraged, however. A failure to find a stable
solution does not mean that the dual string (which we want to become
light at strong coupling and take over the spectrum) does not
exist. It simple means that things are a little more interesting.

It turns out that there is a way~\cite{open} of seeing the fundamental string, 
but only in a particular frame, and the winding circle is a clue as to
how.  Imagine that we place the type~I string theory on a circle in
the $x^1$ direction of radius $R$. We give it $N$ units of momentum in
that direction. Consider the limit where we take $N$ and $R$ to
infinity, to define the string in the ``Infinite Momentum Frame''. In
that frame, the degrees of freedom which survive are those which have
some finite fraction of positive momentum in that direction. So in
the limit, we are probing arbitrarily small distances along that
direction. 

In string theory in this frame, there is a smallest possible ``string bit''
length~\cite{bits} that strings can have. We are therefore able to 
stretch stable strings
of that length, as they cannot break. 
Ultimately, therefore, we see a substructure in that direction
corresponding to being able to resolve a transverse distance ``between''
the individual D9--branes. The D9--branes are point--like along this
special transverse coordinate in this limit.

This is a ``matrix string'' representation of the type~I string in the
infinite momentum frame. The 1+1 dimensional lagrangian for this model
(in the free string limit) is the large $N$ limit of a matrix--valued
Green--Schwarz type~IIB light--cone action with type~I boundary conditions as
one goes around the spatial direction. The size of the matrices is
$N$. Sectors of momentum~$n$ are represented by matrices which have
their boundary conditions twisted by a non--trivial permutation of
$n$ eigenvalues, where $n$ is tuned to be a finite fraction of $N$ as
we take the limit.

There is another language to describe this: to probe those
very short distances, we have T--dualized to a type~IA system, with
$N$ units of winding along the T--dual direction ${\hat x}^1$. The 16
D8--branes are the places where the winding type~IA strings can
end. They are stable, as they cannot break unless another D8--brane is
located at that point. The 1+1 dimensional Green--Schwarz action above controls the
world--volume dynamics of the strings stretching along the light--cone
direction, which is also a static gauge action for the type~IA string.

\begin{figure}[ht]
\hskip3.0cm\psfig{figure=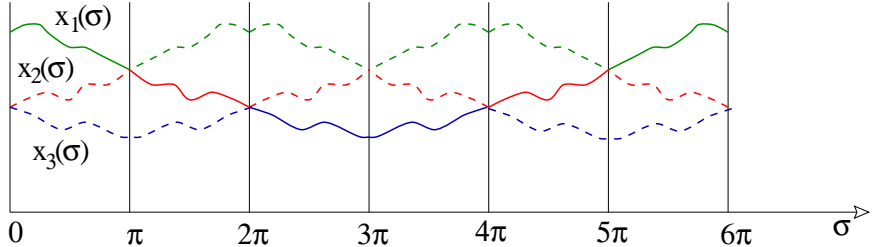,height=1.3in}
\caption{A winding closed unoriented type~I string of length six, made
by acting with an $S_3$ twist with an $\IZ_2$ folded in. This can also
be viewed as an open string of length three and its mirror. The
physical string is the solid line, made by gluing the three functions
together. (Green, then red, then blue, for viewers in colour.) The
strings can end at any multiple of~$\pi$, where there are $SO(16)$
Chan--Paton factors. Odd and even open strings fill out the full
$SO(32)$ gauge multiplet. (See text.)}
\label{unorient}
\end{figure}

This type~IA description of the light--cone matrix type~IB string~\cite{super,open}
can be used to show that the matrix string theory thus
defined has $SO(32)$ gauge symmetry, and that the closed string sector is 
unoriented. 
This latter follows from the crucial fact that one must flip the copy of 
the line interval 
each time one extends to make a multiple cover of the line interval. 
See figure~\ref{unorient}.

We take the $N,R{\to}\infty$ limit and $\alpha^\prime{\to}0$. In supergravity terms,
the resulting limit is the
core geometry of a fundamental string solution of the type~IA
(massive~\cite{gojoe,romans}
type~IIA) supergravity~\cite{super,open}. The string ends on D8--branes at infinity,
which in this limit are domain walls. This solution represents the
fields produced by a type~I string in the limit where we probe to the
substructure showing it stretched between D9--branes.

\section{Type~I String Theory Under the Microscope~II: Instanton Redux}
\subsection{D9--Branes and D5--Branes}
Instead of adding D1--branes to the type~I theory, we can add
D5--branes. Let us add $N$ of them, all initially coincident.  We do
this by adding Dirichlet conditions to the theory requiring that strings
can end at points in the subspace $x^6{-}x^9$, thus defining a 5{+}1
dimensional world--volume of the D5--branes in $x^0{-}x^5$.

In the presence of the brane, the spacetime Lorentz group breaks as
\be
SO(1,9)\supset SO(1,5){\times}SO(4),
\ee where the $SO(1,5)$ refers to the
world--volume fo the D5--brane. The spacetime supercharges decompose
as ${\bf16}{\to}{\bf(4,2,1)}{\oplus}{\bf(4,1,2)}$. Here the {\bf4}'s
are vector representations of the $SO(4)$.  The $SO(1,5)$
representations are given in terms of the $SU(2){\times}SU(2)$ little
group of the Lorentz group.

The D5--branes are annihilated by one of these (choose the first), and
the other generates superpartners in the ${\cal N}{=}1$
supersymmetric model in 5+1 dimensions. This amount of supersymmetry
has an $SU(2)$ R--symmetry which we shall denote as $SU(2)_R$.

We can study the content of the theory in a similar way to the above
discussion for D1--branes, and the following content 
results~\cite{edsmall}: 

The 5--5 strings with coordinates along the world--volume transform as
vectors and give a $USp(2N)$ gauge theory.  The 5--5 strings
transverse give a family of 4 scalars transforming in the
antisymmetric representation of $USp(2N)$. $\Omega$~acts on the
fermions as $\Gamma^1\ldots\Gamma^5$, and correlates their $USp(2N)$
transformation properties with six dimensional chiraity~\cite{douglas}.

There are four DN coordinates ($x^1{-}x^5$), and four DD coordinates
($x^6{-}x^9$) giving the NS sector a zero point energy of
$-{1\over2}{+}{4\over8}{=}0$, with excitations coming from integer
modes in the $1234$ directions, giving initially a four component
boson $h_{\alpha m}^M$ where $M,m$ are $SO(32)$ and $USp(2N)$ indices
respectively. As it acts by exchanging the ends, the $\Omega$
projection relates the 9--5 strings to the 5--9 strings: $(h^M_{\alpha
m})^*=\epsilon_{mn}\epsilon_{\alpha\beta}h_{\beta n}^M$, where the
$\epsilon$'s are the antisymmetric tensors of the $SU(2)_R$ and the
$USp(2N)$. The 5--9 strings therefore give a half--hyper transforming
in the $(2N,32)$ of $USp(2N){\times}SO(32)$.  The R sector also has (as always)
zero point energy 0, with excitations coming from the 6789 directions,
giving a initially four component fermion $\chi$, halved to two
components by $\Omega$, as before.

Returning to our six--dimensional $K3$ compactification for a moment,
we can see~\cite{jhs} how this fits into the anomaly equation~(\ref{anomaly}). 
There, as the transverse part of the space was the
compact $K3$, we had 24 branes and therefore gauge group
$SO(32){\times}USp(48)$ in this point--instanton limit. So
we have $n_V{=}496{+}1176{=}1672$, the dimension of the gauge group. The 9--5 sector
supplies a half--hyper in the $\bf(32,48)$ and the 5--5 sector has a
set of hypers in the antisymmetric of $USp(48)$ which is therefore the
${\bf(1,1128)}{=}{\bf(1,1127)}{+}{\bf(1,1)}$ with the reduction
showing the center of mass position of the mutli--instanton. Together
with the $80$ moduli of $K3$ (equivalent to 20 hypers), this gives
$n_H{=}1916$, as required by the anomaly equation. Once again, further
checking of the anomaly polynomials will reveal that all anomalies
cancel. 

This is of course a special point in the allowed space of vacua, and
we can characterize the classical moduli space as we did previously
for the D1--branes.

\bigskip
\noindent
$\bullet${\it The Coulomb Branch}
\medskip

\noindent
There is a Coulomb branch analogous to that which we found for the
D1--branes, simply corresponding to moving the D5--branes apart. This
is done by giving vevs to the 5--5 hypermultiplets, breaking the
gauge group by the Higgs mechanism. Done in the most complete way, we
get $SU(2)^N$, corresponding to $N$ separated D5--branes~\footnote{This
is a sort of ``non--abelian Coulomb branch'', given that the gauge
group is not some power of $U(1)$.}. The remaining 5--5 hypermulitplet
is a singlet, whose four real components correspond to the position of
the D5--brane in $x^6{-}x^9$. The 9--5 fields on the other hand are
$N$ half--hypers in the $\bf(2,32)$. (It is interesting to veryfy that
this sepctrum is also anomaly free~\cite{jhs}.) The four components of the 9--5
half--hypers correspond to some additional data concerning the
D5--brane, in the $SO(32)$ background which characterizes the ``Higgs
branch''.

\bigskip
\noindent
$\bullet${\it The Higgs Branch}
\medskip

\noindent
At any stage, we can also give vevs to the 9--5 strings. Let us
consider first the simplest case of a single unit, with gauge group
$SU(2)$.  The $\bf(2,32)$ half--hyper allows us to completely Higgs
away the $SU(2)$ fivebrane group and break the $SO(32)$ to
$SO(28){\times}SU(2)$. Let us consider the details of this~\cite{edsmall}.

Denote the half--hypermultplet as $h^M_{\alpha m}$ where $M$ is an
$SO(32)$ index, $m$ is an $SU(2)$ index and $\alpha$ is a doublet
index of the the $SU(2)_R$ symmetry, labelling the components of the
half--hyper. We wish to discover what allowed values of $h^M_{\alpha
m}$ preserve the vanishing of the scalar potential.

The scalar potential can be written as a sum of ``D--terms'' \be
D_{\alpha\beta,mn}{=}\sum_{M}(h_{\alpha m}^Mh_{\beta n}^M+h_{\beta
m}^Mh_{\alpha n}^M). \ee now we may alternatively consider the product
of the gauge $SU(2)$ and the R--symmetry $SU(2)_R$ as $SO(4)$, in
which case we may rewrite $h_{\alpha m}^M$ as a family of 32 $SO(4)$
vectors $h^M_i$ where $i$ is an $SO(4)$ vector index. Alternatively,
$h^M_i$ may be thought of as the components of four vectors, $h_i$, in
a 32 dimensional vector space $V{\approx}\IR^{32}$ on which $SO(32)$ acts.

So the ``D--flatness condition'' (vanishing of the D--terms) is
equivalent to \be (h_i,h_j)=\rho^2\delta_{ij} \ee where
$(\phantom{h},\phantom{h})$ is the inner product on $V$. So the
vectors $e^i{=}h^i/\rho$ are orthonormal vectors. Choosing the four
orthonormal $e^i$ in $V$ breaks the $SO(32)$ down to
$SO(28){\times}SU(2)$, where we get the extra $SU(2)$ by treating
equivalent $h^i$'s which are gauge related by $SU(2)$. Therefore that
$SU(2)$ arises as a result of dividing by the original fivebrane gauge
group to get the correct moduli space of vacua.

So what do we have? The allowed values of $h$ leave us with gauge
group $SO(28){\times}SU(2)$. The subgroup of $SO(32)$ which commuted
with this gauge group is an $SU(2)$, which is fully specified by
choosing four parameters, the scale $\rho$ and the orientation of the
$e^i$ basis. This is exactly the data needed to specify an $SO(32)$
instanton gauge bundle with structure group $SU(2)$. The scale size of
the instanton (``thickness'' of the fivebrane) is $\rho$.

What we have done is described the one--instanton version of the
``hyperK\"ahler quotient'' technique~\cite{quotient} of constructing instantons. 
In the general case of $USp(2N)$ with half hypers in the $(2N,32)$, we get
the full ADHM construction of $N$ $SO(32)$ instantons~\cite{edsmall,adhm}.
So we discover that the Higgs branch of our gauge theory is
parameterized by the 9--5 fields and is isomorphic to the moduli space
of $SO(32)$ instantons~\footnote{This immediately
generalizes. The important features in the above discusssion was the
type of half--hypermultiplet which appeared, in a bi--fundamental
representation of a product gauge group where the products came from
one brane inside another brane. such a multiplet will always appear
when there are four DN direction which happens when a D$p$--brane is
inside a D$(p{+}4)$--brane. Therefore, a D1--brane is an instanton of a
D5--brane, a D0--brane is and instanton of a D4--brane, and so 
on~\cite{douglas}.

Yet another way to see this is from the world--volume action of the
D$(p{+}4)$--brane. It contains Chern--Simons couplings of the form
$\int\!A_{p+1}{\wedge}F{\wedge}F$ where $A_{p{+}1}$ is a R--R field,
and the integral is over the $(p{+}5)$--dimensional
worldvolume. Therefore, a D$p$--brane acts as a source of $F{\wedge}F$
in the world volume by virtue of being a source of $A_{p{+}1}$.
It is also worth noting that there is also a term of the form 
$\int\!A_{p+1}{\wedge}R{\wedge}R$, which means that wrapping the
D$(p{+}4)$--brane on a space of non--zero $R{\wedge}R$, like $K3$, will
result in it appearing as a $p$--dimensional source of $A_{p{+}1}$
R--R charge.}.

This harmonizes perfectly with our discussion of section~3. The supergravity
and anomaly analysis led us to anticipate a new massless sector of the theory,
arising as the vev of a hypermultiplet goes to zero. We see that this massless
sector is in the form of the gauge theory on the world--volume of
type~I D5--branes. Let us  call this D5--brane
gauge theory with this particular content the ``ADHM gauge theory'' for short, after 
the structure of its Higgs branch analyzed above.

\subsection{Cadenza: ADHM Gauge Theory as String Theory on a Throat}
The $5{+}1$ dimensional ADHM gauge theory 
has a well defined Coulomb and Higgs branch. At weak coupling (the infra--red (IR)),
its Coulomb branch supplies  the missing massless 
degrees of freedom when type~IB the supergravity description of instanton fivebranes 
break down, as we saw.

It is interesting to speculate that the heterotic throat description which we were led to
earlier did not
break down so much as simply take us to a realm where we were unsure
of the interpretation of some of our tools. There is the possiblilty
that the D5--brane description and the throat description might be
complementary~\cite{mythroat}, which is the oft--repeated lesson of duality. The
linear dilaton description, with its exact (WZNW+FF) conformal field
theory (CFT) representation, might still capture the physics of the
gauge theory after all. Previously, however, we said that the
heterotic string perturbation theory ---including conformal field
theory--- cannot describe the D5--brane physics, not the least because
the allowed gauge groups ({\it e.g.} as large as $USp(48)$) in
our~$K3$ example above) would give a central charge much greater than
24, so surely this is a contradiction?

It is not a contradiction~\cite{mythroat} for at least two reasons:

\begin{itemize}
\item The conformal field theory of the
throat is {\it completely disconnected} from the theory outside the
throat. Indeed, it has been long known that throat conformal field
theories are notoriously difficult to connect to the asymptotically
flat region: The operators which describe the widening of the throat
are singular~\cite{mouth}. Now we know why. The correct interpretation of the
string theory down the throat, as captured for example by the exact
(WZW+FF) CFT is that it is a dual description of the theory on the
world volume of the D5--branes. 

\item The gauge theory does not need to be explicit. This certainly has to be
true, for the reasons stated above, and so it must be encoded in a
different way. Given that the CFT naively seems to be describing a strongly
coupled heterotic background, which invalidates many of the standard
interpretations, is clear that there is some room for a new
interpretation.
\end{itemize}

In capturing the physics of the ADHM gauge theory in a dual model, it
is crucial to realize that we only need find the {\it gauge invariant}
physics. This is why we need not find the gauge particles explicitly
down the throat. An example of the gauge invariant information which
the dual representation should capture is the moduli space of
vacua. Part of this is the Coulomb branch representing the patterns of
Higgsing which can occur as the D5--branes are moved around in
$x^6{-}x^9$ directions, including possible coincidences. This is
something which should be captured in the throat limit. Indeed,
there are arguments to support this, with the $USp(2N)$ structure showing up in
the modular invariant used to build the CFT partition function. This
controls the moduli space of marginal deformations of the CFT which is
isomorphic to the Coulomb branch of the ADHM gauge theory.
This can be thought of as another generalization of the AdS/CFT
correspondence, where gauge theory and geometry complement one
another, this time with a non--trivial dilaton playing a crucial
role~\footnote{This relationship was pointed out in paper ~\cite{mythroat}, but the
author carelessly did not put the word ``holography'' anywhere in the
title, abstract or body of the paper. Fashion--conscious readers
should therefore instead see ref.~\cite{oferetal} for later (independent)
 work on holography and linear dilatons, in the context of the
type~IIA NS--NS branes. That work also discusses in detail how the holographic
correspondence should work, given the peculiar properties of the throat theory.} 
instead of negative cosmological constant~\cite{myers}.

Notice that the string coupling is $g_s{=}\e{-\sigma/R}$.
The region where the throat conformal field theory does have the traditional
interpretation is when $\sigma{\to}{+}\infty$, which is a {\it continuation} of the throat
(continue the left--hand part of the diagram~\ref{fig:throat}
infinitely to the right). While staying down the throat, we have 
continued to a region where the heterotic string is weakly coupled. What does this limit
correspond to for the putative dual gauge theory?

As a six dimensional gauge theory, heading towards the ultra--violet (UV), the theory
should break down at (mass)$^2$ scale $1/g^2_{YM}{=}1/(\alpha^\prime g_s)$, where $g_s$ 
is the type~I string coupling, which is going strong if we keep $\alpha^\prime$ fixed
for a moment. As $g_s{\to}\infty$, we can 
ask what this physics looks like in the heterotic picture. We see that the (mass)$^2$
scale is simply $1/\alpha^\prime$ in heterotic string units --- the heterotic
string coupling has gone from the formula. 
The independence of the gauge coupling of the string coupling in heterotic variables is
a clue that there is a sensible theory~\cite{seibergii} \footnote{This type of
theory, associated with the four types of NS--NS brane in the 
$SO(32)$, $E_8{\times}E_8$, type~IIA and type~IIB string theories, is sometimes called a 
``little string'' theory, or a ``micro--string'' theory~\cite{microverlinde}. 
See the lectures of H.~Verlinde at this school.}
living on the world--volume of the NS--NS brane
in the limit of vanishing heterotic string coupling. 
In other words, the heterotic string is telling us
that there is no real problem with the ADHM theory in the UV, and we have found
a description in heterotic variables in terms of the throat CFT! Indeed, we can control the
approact to the UV in terms of the usual small $\alpha^\prime$ expansion.

So we indeed have a new gauge theory/geometry correspondence. The theories are complementary,
as the weakly coupled (IR) limit of the ADHM gauge theory is best described in terms
of the defining lagrangian, because the throat theory is strongly coupled, while the UV
of the ADHM theory is best described in terms of the weakly coupled string theory
propagating on the throat, described by the WZNW+FF
conformal field theory.

It is worth noting another attractive feature of this possibility. To
properly describe string theory propagating on the throat in the
weakly coupled supergravity limit, we should really take $N$ large as
$\alpha^\prime{\to}0$, in order to keep the radii ($R{=}\sqrt{N\alpha^\prime}$) of the
$S^3$'s large enough to keep the curvature corrections down.  
We actually seem to be able to describe much more than this restriction would suggest.
 The exact CFT
representation means that we have not only the leading order in
$\alpha^\prime$ description of the geometry, but the full
$\alpha^\prime$ description. Furthermore note~\cite{chs} that the precise
combination of the $SU(2)_N$ WZNW model and the Feigin--Fuchs theory
is such that the value of $N$ is cancelled exactly in the central
charge formula: The WZNW model and the three free fermions needed for
supersymmetry~\footnote{Note that chiral rotation~\cite{robin} needed to make the
supersymmetry fermions free~\cite{rohm} has the effect of shifting the level $N$
to $N{-}2$.} give $c{=}3{-}{6\over N}{+}{3\over2}$, while the
Feigin--Fuchs scalar plus its fermion gives $c{=}1{+}{6\over
N}{+}{1\over2}$. Their total central charge is exactly $6$, which
together with the $c{=}9$ from the superstring propagating on the flat
$x^0{-}x^5$ gives the required 
$c_{\rm tot}{=}15$~\footnote{This is for the right hand, supersymmetric,
side of the theory. The left hand side has $c_{\rm tot}{=}26$, which can
be arranged~\cite{mythroat}.} This exact
formula for the central charge should be reflected in the properties
of the fields and vertex operators in the full conformal field
theory. Therefore, when endowed with the correct interpretation, the
exact CFT should contain the complete story for all $N$ and all~$\alpha^\prime.$

Therefore the full ``holographic'' statement is that the ADHM gauge theory at
any $N$ is described by the heterotic string propagating on the throat geometry,
which is succinctly given by the exact WZWN+FF conformal field theory. 

It is also natural to expect that a gauge theory interpretation of
this type will exist for the many other types of exact throat CFT's in
existence in the literature.

\section{Recapitulation}
Duality and D--branes have taught us 
a number of lessons about non--perturbative string theory. 
Let us list a few of them. First, duality says that:
\begin{itemize}
\item{The low--energy effective actions of the massless fields in the string spectra
map into one another  under the
duality transformation.}
\item{The solutions of the corresponding equations of motion also map into each other. 
In particular:
\begin{itemize}
\item{The NS--NS charged fundamental string solution, light at weak coupling
and representing the perturbative string,
maps into a R--R charged solitonic string, heavy at weak coupling 
(${\rm tension}{\sim}1/g_s$)
in the dual string theory.}
\item{The NS--NS charged fivebrane soliton solution (tension ${\sim}1/g^2_s$) 
maps into a special type of R--R  charged fivebrane soliton, (tension~${\sim}1/g_s$).
Both fivebranes are $SO(32)$ instantons. The scale size of the instanton is equivalent 
to the thickness of the fivebrane.}
\end{itemize}}
\item{As supergravity solutions, the various fivebrane solitons have special properties 
as the instanton size vanishes:
\begin{itemize}
\item{The NS--NS fivebrane has an infinite throat at its core, down which the
string coupling grows exponentially. Meanwhile, the R--R fivebrane is singular at the core, 
although the string coupling is weak there. }
\end{itemize}
So the supergravity description evidently has problems when the fivebranes 
are thin.}
\end{itemize}
We need another description of
the physics in the small instanton/thin five--brane limit. This is where D--branes come in:
\begin{itemize}
\item{The type~I supergravity is singular because we missed some massless states.
There are massless vectors associated with an enhanced gauge symmetry which appears
in the small instanton limit.}
\item{The small instanton (thin fivebrane) has a description as a D5--brane, introduced
 into
type~I conformal field theory by adding Dirichlet boundary conditions. 
$N$~D5--branes in type~I
have  a $USp(2N)$ gauge theory with two classes of hypermultiplets possessing transformation 
properties and couplings which constitute what we called an ``ADHM gauge theory''. 
The space of 
allowed vevs of some of the hypermultiplets is isomorphic
to the moduli space of instantons. 
The hypermultiplet vev which controls the thickening of the fivebrane also 
gives masses to the vectors, taking us back to the supergravity description.}
\end{itemize}

So we saw that gauge theory supplements the type~I supergravity description. 
It is hasty to throw out the heterotic supergravity throat description, however.
It gives a dual representation of the Coulomb branch physics of the ADHM gauge theory. 
In particular, it has an exact conformal field theory description, with a strongly coupled
regime (hard to interpret) and, by continuation, a weakly coupled regime, which supplies
a complementary description of the UV of the ADHM theory.
Indeed, the gauge invariant information
---{\it e.g.,} the moduli space of deformations--- can be encoded in the heterotic 
modular invariant of the conformal field theory~\cite{mythroat}. 

This is another type of holography, this time with the linear
dilaton playing the role that negative cosmological constant  did in the AdS 
case~\cite{myers,oferetal}.
Notice that although the throat was properly a supergravity solution in the large $N$,
small $\alpha^\prime$ limit, the form of the exact conformal field theory
description suggests that this 
is actually true for all $N,\alpha^\prime$ which is at the least, very interesting. We have 
therefore the complete holographic statement that the 
ADHM gauge theory is dual to heterotic string theory propagating on a throat, described
by an exact conformal field theory.

\medskip
\begin{center}
* * *
\end{center}
\medskip

So we have come full circle over the last ten years. We started out with 
 large $N$ matrix model descriptions of 
very low dimensional string theory. The double scaling limit 
allowed for a complete description
at all orders in the $1/N$ expansion (which is isomorphic to the string genus
expansion), and non--perturbative information as well. The non--perturbative effects
were associated with the tunneling of a matrix eigenvalue, giving $\e{-N}{\sim}\e{-1/g_s}$
effects.

D--branes are responsible for $\e{-1/g_s}$ effects in critical string theory. They
have a gauge theory on their world volume. These gauge theories are dual to
string theories in a manner superficially similar to the simpler matrix models. 
It is hopefully clear, after
the discussion in the five studies collected here, that these things are 
not coincidences: The gauge theories are also ``matrix models'' in an enlarged sense, and 
D--branes
are eigenvalues in this framework. Clearly, the eigenvalues of the simpler
matrix models of ten years ago correspond to D--branes of the low dimensional 
``non--critical'' string theories, but whether this is a
 useful concept is not clear to the author.

A closer
examination shows that the dual string theories to (at least) some of the gauge theories
are string theories propagating on background spacetimes with unusual properties.  
In the simplest matrix models, we were able to solve the string theories
exactly. In the case of AdS backgrounds, 
the correspondence is tested mainly at string tree level, which is supergravity. 
For the linear dilaton
background however, it appears that the full stringy correspondence might be captured 
by an exact CFT.

Duality, in its various forms clearly has much to teach us about the nature of 
fundamental physics. It has
pulled together a number apparently discordant approaches and  hints
over the years into single harmonious narrative. 
Matrix models, gauge theory, D--branes
and other extended objects have been the chief means of instruction so far. 
With surprises happening with astonishing regularity in the field, the only 
safe prediction
is that there is much more excitement in store for us in the next ten years.

\bigskip
\hskip3.5cm\psfig{figure=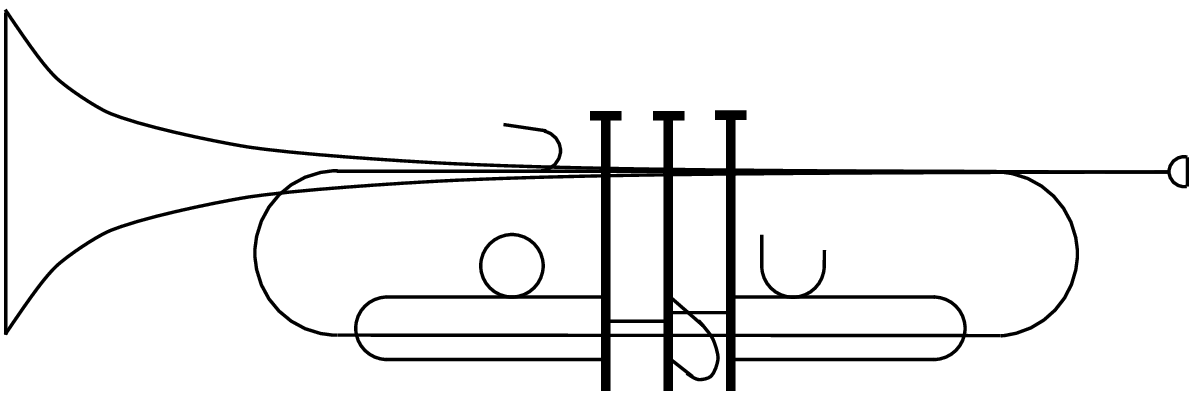,height=0.5in}

\section*{Acknowledgments}
This work was supported by an NSF CAREER grant, \#9733173. I am
grateful to the organisers of the 1998 Trieste Spring School for the
invitation to give these lectures, and to them and the staff at the
Abdus Salam Centre for Theoretical Physics for helping to make my stay
there such a pleasant one. Many thanks to S. J. Butler
for patience and 
hospitality while the crucial finishing touches were applied to this work.

\end{document}